\documentclass[aps, prd, preprint, a4paper, 12pt, superscriptaddress, nofootinbib]{revtex4-1}
\usepackage[dvipdfmx]{graphicx}
\usepackage{amsmath,amssymb}
\usepackage{comment}
\usepackage{here}
\usepackage{docmute}
\usepackage{simplewick}
\usepackage{layout}
\usepackage[margin=2.4cm]{geometry}
\usepackage{color}

\usepackage[normalem]{ulem}
\usepackage{cancel}

\begin{document}

\title{\bf Interaction potentials for two-particle states with non-zero total momenta in lattice QCD}

\author{Yutaro Akahoshi}
\altaffiliation{Present address: Quantum Laboratory, Fujitsu Research, Fujitsu Limited. Kanagawa 211-8588, Japan}
\affiliation{Center for Gravitational Physics, Yukawa Institute for Theoretical Physics\\Kyoto University, Kyoto 606-8502, Japan}
\affiliation{RIKEN Nishina Center (RNC), Saitama 351-0198, Japan}

\author{Sinya Aoki}
\affiliation{Center for Gravitational Physics, Yukawa Institute for Theoretical Physics\\Kyoto University, Kyoto 606-8502, Japan}
\affiliation{RIKEN Nishina Center (RNC), Saitama 351-0198, Japan}

\preprint{YITP-22-124} \preprint{RIKEN-QHP-516}
%%%%% abstract %%%%%
\begin{abstract}
In this study, we extend the HAL QCD method to a case where a total momentum of a two-particle system is non-zero and apply it to the $I=2$ S-wave $\pi\pi$ scattering in order to confirm its validity.
We derive a fundamental relation of an energy-independent non-local potential defined in the center of mass frame
with NBS wave functions in a laboratory frame.
Based on the relation, we propose the time-dependent method to extract potentials,
often used in practice for the HALQCD method in the center of mass frame.
For numerical simulations in the $I=2$ $\pi\pi$ system, we employ (2+1)-flavor gauge configurations on a $32^3 \times 64$ lattice at the lattice spacing $a \approx 0.0907$ fm and $m_{\pi} \approx 700$ MeV.
Both effective leading order (LO) potentials and corresponding phase shifts obtained in laboratory frames agree with
those obtained in the center-of-mass frame by the conventional HAL QCD method within somewhat larger statistical errors.
In addition, we observe a consistency in scattering phase shifts between ours and results by the finite-volume method as well.
The HAL QCD method with non-zero total momenta, established in this study, brings more flexibility to the HAL QCD method,
which enables us to handle systems having the same quantum numbers with a vacuum or to access energy regions prohibited in the center of mass frame.

\end{abstract}
%%%%% abstract end %%%%%
\maketitle

%%%%% introduction %%%%%
\section{Introduction}\label{sect:intro}

A first-principle determination of hadron interactions in quantum chromodynamics (QCD) is
one of the most important challenges for understanding
the nature of hadronic resonances observed in experiments.
In recent years, hadron interactions have been actively studied in lattice QCD
by employing
two methods: the finite-volume method~\cite{Luscher:1990ux,Rummukainen:1995vs,Hansen:2012tf} and
the HAL QCD method~\cite{Ishii:2006ec,Aoki:2009ji,Aoki:2011ep,HALQCD:2012aa}.
Theoretically, both methods rely on the Nambu-Bethe-Salpeter (NBS) wave function, which links to the scattering S-matrix in QCD.
The finite-volume method gives a formula relating finite-volume energy spectra to scattering phase shifts in the infinite volume,
by considering behaviors of NBS wave functions in an asymptotic region.
In the HAL QCD method, on the other hand, one directly extracts energy-independent non-local potentials through
spatial dependences of NBS wave functions in an interacting region,
and then calculates scattering phase shifts by solving Schr\"odinger equations with these potentials in the infinite-volume.
Armed with the time-dependent method~\cite{HALQCD:2012aa} and the multi-channel extension~\cite{Aoki:2011gt}, the HAL QCD method has been successfully applied to two-baryon systems with various pion masses (see Ref.~\cite{Aoki:2020bew} and references therein for the recent status).
Recently resonance studies such as the $\rho$ meson have become possible thanks to rapid improvements in calculation techniques of all-to-all quark propagators~\cite{Kawai:2017goq,Kawai:2018hem,Akahoshi:2020ojo,Akahoshi:2019klc}.

In the finite-volume method,
not only the center of mass frame but also laboratory frames with non-zero total momenta~\cite{Rummukainen:1995vs}
are employed for calculations  of two hadron spectra,
in order to access different finite-volume energies as many as possible for a given volume,
which provide  energy dependences of scattering observables precise enough for determinations of resonance parameters~\cite{Briceno:2017max}.

In the HAL QCD method, on the other hand,
all calculations so far have been made in the center of mass frame,
since the time-dependent HAL QCD method~\cite{HALQCD:2012aa} does not require isolation of an energy eigenstate but
utilizes all elastic states below inelastic thresholds at some degree.
The recent $\rho$ resonance study, however, reveals that
P-wave scattering phase shifts can not be determined precisely in a low-energy region not accessible in the center-of-mass frame
due to non-zero relative momenta of the P-wave~\cite{Akahoshi:2021sxc}.
Thus the HAL QCD method in laboratory frames with non-zero momenta will provide a better alternative for
a more precise determination of P-wave scatterings.
In addition, the HAL QCD method in laboratory frames is mandatory to
investigate hadron resonances having the same quantum numbers with a vacuum such as
the $I=0$ S-wave $\pi\pi$ interaction corresponding to the $\sigma$ resonance,
since a large mixing to a vacuum state in the center of mass frame is much reduced in a laboratory frame.

An extraction of HAL QCD potentials using laboratory frames with non-zero total momenta has been theoretically formulated and
briefly reported in~\cite{Aoki:2020cwk}.
Applications of the method to hadron systems in lattice QCD, however, require
numerically demanding calculations of all-to-all quark propagators,
which are employed to construct appropriate source operators with non-zero total momenta.
As already mentioned, recent algorithmic improvements make it possible to calculate all-to-all propagators
with reasonable numerical costs in the HAL QCD method.
We thus apply the method of HAL QCD potentials in laboratory frames to
one of the simplest  two hadron systems, the $I=2$ S-wave $\pi\pi$ system at  $m_{\pi} \approx 700$ MeV,
in order to see how the theoretical formulation works in actual lattice QCD simulations.
This system does not have quark annihilation diagrams, and its interaction has been already known
to be repulsive at low energy~\cite{Sasaki:2013vxa,Kawai:2017goq,Akahoshi:2019klc}.
Employing a time-dependent method reformulated for correlation functions in laboratory frames,
we extract effective leading order (LO) potentials successfully from correlation functions with total momenta $|{\bf P}| = 2\pi/L$ and $4\pi/L$ for the first time.
We then calculate physical observables such as scattering phase shifts using potentials obtained in laboratory frames,
which are compared with results obtained  not only from the conventional HAL QCD potential in the center of mass frame
but also by the finite volume method through finite volume spectra.
We confirm a consistency among all results, which validates our new method to extract HAL QCD potentials with non-zero total momenta
both theoretically and numerically.

This paper is organized as follows. In Sect.~\ref{sect:potential},
we formulate a procedure to extract HAL QCD potentials from correlation functions in laboratory frames, using
a transformation property of NBS wave functions under Lorentz transformation.
In Sect.~\ref{sect:demonstration}, we present numerical results on potentials and scattering phase shifts in the $I=2$ $\pi\pi$ system.
We give summary and discussion in Sect.~\ref{sect:summary}.
A detailed analysis on systematic errors associated with laboratory frame calculations is given in Appendix~\ref{appx:lab_hal_sys}.
Preliminary results in this study have been already reported in the conference proceedings\cite{Aoki:2021tcu}.

%%%%% introduction end %%%%%

%%%%% method %%%%%
\section{HAL QCD potential from the NBS wave function in the laboratory frame} \label{sect:potential}
\subsection{Lorentz transformation of the NBS wave function} \label{subsect:Lorentztrsf}
Let us consider two scalar particles with same mass $m$ in Minkowski spacetime.
The corresponding NBS wave function in a general frame of reference is defined as
\begin{equation} \label{eq:NBS}
  \psi_{k_1,k_2} (x_1, x_2) = \langle 0 | {\rm T} \{ \phi(x_1) \phi(x_2) \} | k_1, k_2 \rangle,
\end{equation}
where $\phi(x)$ is a scalar field operator and $| k_1, k_2 \rangle$ is an asymptotic in--state of two particles with four--momenta
$k_1$ and $k_2$. Explicitly $k_i =\left(\sqrt{{\bf k_i}^2 +m^2}, {\bf k}_i\right)$ for $i=1,2$.
Lorentz transformation acts on field operators and asymptotic states as
\begin{eqnarray}
  U(\Lambda) \phi (x) U^{-1}(\Lambda) = \phi(x'), \label{eq:trsf_op} \\
  U(\Lambda) | k_1, k_2 \rangle = | k'_1, k'_2 \rangle, \label{eq:trsf_state}
\end{eqnarray}
where $U(\Lambda)$ is a unitary operator which implements Lorentz transformation on the states and prime symbols represent transformed objects, for example, $x'^{\mu} = {\Lambda^{\mu}}_{\nu} x^{\nu}$.
%Eq.~(\ref{eq:trsf_state}) defines the normalization and phase of the asymptotic state.
Using eqs. (\ref{eq:NBS}), (\ref{eq:trsf_op}) and (\ref{eq:trsf_state}), we can derive a relation between two NBS wave functions in different frames as
\begin{equation} \label{eq:trsf_NBS}
  \psi_{k_1,k_2}(x_1,x_2) = \psi_{k'_1, k'_2}(x'_1, x'_2).
\end{equation}
Furthermore,  a relation $\phi(x) = e^{i \hat P \cdot x} \phi(0) e^{-i P \cdot x}$, where $\hat P$ is an energy-momentum operator and $\hat P \cdot x = \eta_{\mu \nu} \hat P^{\mu} x^{\nu}$, implies
that the NBS wave function is factorized into a center--of--mass plane wave and a relative wave function as
\begin{equation} \label{eq:factorized_NBS}
  \psi_{k_1, k_2} (x_1, x_2) = \varphi_{k_1,k_2}(x) e^{-i W X^0 + i {\bf P} \cdot {\bf X} },
\end{equation}
where $W=\sqrt{{\bf k}_1^2+m^2} + \sqrt{{\bf k}_2^2+m^2}$ and ${\bf P} = {\bf k}_1 + {\bf k}_2$ are total energy and momentum, respectively.
A center--of--mass coordinate $X$ and a relative coordinate $x$ are defined by
\begin{eqnarray}
  X &=& \frac{x_1 + x_2}{2}, \quad
  x = x_1 - x_2.
\end{eqnarray}
Since $\eta_{\mu \nu} P^{\mu} X^{\nu} = W X^0 - {\bf P} \cdot {\bf X}$ is Lorentz invariant, eqs.~(\ref{eq:trsf_NBS}) and (\ref{eq:factorized_NBS}) give a relation between relative wave functions in different frames as
\begin{equation} \label{eq:trsf_relNBS}
  \varphi_{k_1,k_2}(x) = \varphi_{k'_1,k'_2}(x').
\end{equation}
%Eq. (\ref{eq:trsf_relNBS}) plays a fundamental role in our formulation
%since it enables us to calculate potentials defined in Center--of--mass (CM) frame by using the NBS wave function in Laboratory frame.

The HAL QCD potential is defined in the center of mass (CM) frame,
whose total energy and  momentum satisfy
\begin{eqnarray}
W^* = \gamma ( W - {\bf v} {\bf P} ), \quad {\bf P}^* = \gamma({\bf P} -{\bf v} W)=0,
\end{eqnarray}
where quantities with and without $*$ refer to those in the  CM and laboratory (Lab) frames, respectively, and
a boost factor $\gamma$ is defined by $\gamma={1\over \sqrt{1-{\bf v}^2}}$.
The CM condition ${\bf P}^* = 0$ implies
\begin{eqnarray}
{\bf v} ={\bf P}/W, \quad (W^*)^2 = W^2 - {\bf P}^2,  \quad \gamma ={W\over W^*}.
\end{eqnarray}
With these ${\bf v}$ and $\gamma$,
relative coordinates in CM and Lab frames are related as
\begin{eqnarray}
  x^{*0} &=& \gamma (x^0 - {\bf v} \cdot {\bf x}_{\parallel}), \quad
  {\bf x}_{\parallel}^{*} = \gamma ({\bf x}_{\parallel} - {\bf v} x^0), \quad
  {\bf x}_{\perp}^{*} = {\bf x}_{\perp},
 \label{eq:tranM}
\end{eqnarray}
where $\parallel$ and $\perp$ mean vectors parallel and perpendicular to ${\bf v}$, respectively.

\subsection{HAL QCD potential from the NBS wave function in the laboratory frame} \label{subsect:HALpotential}
We now move on to Euclidean spacetime, in which actual lattice simulations are carried out.
In  Euclidean coordinates, eq.~(\ref{eq:tranM}) reads
\begin{eqnarray} \label{eq:trsf_x_eucl}
  x^{*4} &=& \gamma (x^4 - i {\bf v} \cdot {\bf x}_{\parallel}), \quad
  {\bf x}_{\parallel}^{*} = \gamma ({\bf x}_{\parallel} + i {\bf v} x^4), \quad
  {\bf x}_{\perp}^{*} = {\bf x}_{\perp}.
\end{eqnarray}
In the CM frame, the relative NBS wave function at fixed $x_4^*$ satisfies the Helmholtz equation at large separation as
\begin{equation}
  (\nabla^{*2} + k^{*2}) \varphi_{k^*_1,k^*_2}({\bf x}^*,x^{*4}) = 0 \qquad (r^* = |{\bf x}^*| > R),
\end{equation}
where $R$ is an interaction range and  $k^{*} = \vert {\bf k}^*\vert $ for a relative momentum
${\bf k}^*={\bf k}^*_1 = - {\bf k}^*_2 $ in the CM frame,
and its radial part with an angular momentum $l$ behaves as~\cite{Aoki:2009ji,Aoki:2013cra}
\begin{equation}
  \varphi^l_{k^*_1,k^*_2}(r^*,x^{*4}) \approx A_l(x^{*4},k^*) e^{i \delta_l(k^*)} \frac{\sin \left( k^* r^* - l \pi / 2 + \delta_l(k^*) \right)}{k^* r^*},
\end{equation}
where $A_l(x^{*4},k^*)$ is an overall factor and $\delta_l$ is the scattering phase shift, which is equal to the phase of S-matrix in the scalar field theory.
%As in the case of the equal time NBS wave function ($x_4^*=0$) employed in previous studies\cite{xxx},
We construct an energy--independent non--local potential through the Schr\"odinger--type equation as
\begin{equation} \label{eq:Scheq}
  \frac{1}{2 \mu}(\nabla^{*2} + k^{*2}) \varphi_{k^*_1,k^*_2}({\bf x}^*,x^{*4}) = \int d^3{\bf y^{*}} U_{x^{*4}}({\bf x^*},{\bf y^*}) \varphi_{k^*_1,k^*_2}({\bf y}^*,x^{*4}),
\end{equation}
where $\mu = m / 2$ is the reduced mass.
A subscription $x^{*4}$ of $U$ represents a scheme of the potential that $U$ is defined from NBS wave functions at a relative time separation $x^{*4}$.
In general, a potential depends on a choice of hadron operators (relative time separation, smearing, etc.) in the NBS wave function and it is referred to as the scheme dependence~\cite{Kawai:2017goq,Aoki:2012tk}.
Physical observables extracted from potentials in different schemes, of course, agree with each other by construction.
In practice, we introduce the derivative expansion to treat the non-locality of the potential as
\begin{equation} \label{eq:derivexp}
  U_{x^{*4}}({\bf x^*},{\bf y^*}) = \sum_i V^{i}_{x^{*4}}({\bf x^*}) \left(\nabla^{*2}\right)^{i} \delta({\bf x^*}-{\bf y^*}),
\end{equation}
where $V^i_{x^{*4}}({\bf x^*})$ are local coefficient functions in the expansion
\footnote{We do not include terms with odd powers of ${\bf \nabla}$ here. This choice can be also regarded as a scheme of the potential.
}.
Thus the effective leading-order (LO) potential is simply given by
\begin{equation}
  V^{\rm LO}_{x^{*4}}({\bf x}^*) = \frac{(\nabla^{*2} + k^{*2}) \varphi_{k^*_1,k^*_2}({\bf x}^*,x^{*4})}{2 \mu \varphi_{k^*_1,k^*_2}({\bf x}^*,x^{*4})}.
\end{equation}

Now we are ready to construct an interaction potential from the NBS wave function in the Lab frame.
According to  eqs.(\ref{eq:trsf_relNBS}), (\ref{eq:trsf_x_eucl}) and (\ref{eq:derivexp}), the relative NBS wave function in the Lab frame satisfies
\begin{equation} \label{eq:LabHAL_pot}
  \begin{split}
    \frac{1}{2 \mu}&(\nabla_{\perp}^{2} + \gamma^2 (\nabla_{\parallel} + i {\bf v} \partial_{x^4})^2 + k^{*2})
     \varphi_{k_1,k_2}({\bf x},x^4) \\
     &= \sum_i V^{i}_{\gamma(x^4 - i {\bf v} \cdot {\bf x}_{\parallel})} \left({\bf x}_{\perp},\gamma({\bf x}_{\parallel} + i {\bf v} x^4) \right) \left(\nabla_{\perp}^2+\gamma^2(\nabla_{\parallel} + i {\bf v} \partial_{x^4})^2 \right)^{i} \varphi_{k_1,k_2}({\bf x},x^4).
  \end{split}
\end{equation}
To extract a meaningful potential from this equation, we have to set $x^4 = 0$, since ${\bf x}^*_{\parallel}$ becomes complex with non-zero $x^4$.
We also need to fix ${\bf x}_{\parallel}$ in order to specify $x^{*4}$,
the scheme of the potential, since $x^{*4}$ depends on ${\bf x}_{\parallel}$. In this paper, choosing ${\bf x}_{\parallel} = 0$, we extract
a potential in the equal--time scheme.
As a result, the LO potential in the equal--time scheme is given by
\begin{equation} \label{eq:LabHAL_LOpot}
  V^{\rm LO}_{x^{*4}=0}({\bf x}^*_{\perp}) = \left. \frac{(\nabla_{\perp}^{2} + \gamma^2 (\nabla_{\parallel} + i {\bf v} \partial_{x^4})^2 + k^{*2}) \varphi_{k_1,k_2}({\bf x},x^4)}{2 \mu \varphi_{k_1,k_2}({\bf x},x^4)} \right|_{x^4 = 0, {\bf x}_{\parallel} = 0},
\end{equation}
where we set $x^4 = 0$ and ${\bf x}_{\parallel} = 0$ after taking derivatives in the right--hand side.

In lattice simulations, we put a system in a box of size $L \times L \times L$ with periodic boundary conditions in the Lab frame.
We define a correlation function of the target two-hadron system as
\begin{equation} \label{eq:latNBS}
  F_{\phi\phi, {\bf P}}(x_1, x_2) = \langle {\rm T} \phi(x_1) \phi(x_2) {\mathcal J}_{\phi \phi}({\bf P},0) \rangle,
\end{equation}
where ${\mathcal J}_{\phi \phi}({\bf P},0)$ creates two-particle states with total momentum ${\bf P}$ at $X^4=0$, which is quantized as ${\bf P} = \frac{2 \pi}{L} {\bf n}_{\rm total} \ ({\bf n}_{\rm total} \in {\bf Z}^3)$.
This correlation function can be written as
\begin{eqnarray}
    F_{\phi\phi, {\bf P}}(x_1, x_2) &=& e^{i {\bf P \cdot X}} \sum_{\bf n} B_{\bf n} \varphi_{W_{\bf n}}(x) e^{-W_{\bf n} X^4} + ({\rm inelastic \ contributions})
     \label{eq:latNBS_exp0} \\
    & \to& e^{i {\bf P \cdot X}} B_{\rm min} \varphi_{W_{\rm min}}(x) e^{-W_{\rm min} X^4}, \qquad (X^4 \gg 1),
 \label{eq:latNBS_exp}
\end{eqnarray}
where
\begin{eqnarray}
W_{\bf n} &=& \sqrt{{\bf k}_1+m^2} + \sqrt{{\bf k}_2+m^2}, \quad
{\bf k_1} + {\bf k_2} = {\bf P} ,\\
B_{\bf n} &=& \langle  k_1,k_2 \vert {\mathcal J}_{\phi \phi}({\bf P},0) \vert 0\rangle, \quad k_1^0=\sqrt{{\bf k}_1+m^2},\
k_2^0= \sqrt{{\bf k}_2+m^2},
\end{eqnarray}
$W_{\rm min}$ is the minimum value among $W_{\bf n}$, and  corresponding $B_{\rm n}$ and $\varphi_{\rm n}$ denote $B_{\rm min}$ and $\varphi_{\rm min}$, respectively.
 Therefore, we can extract the NBS wave function of the lowest energy state through this correlation function at a large CM time $X^4$.
Note that these relative NBS wave functions have a periodicity depending on ${\bf P}=\frac{2 \pi}{L} {\bf n}_{\rm total}$ as
\begin{equation} \label{eq:d-periodicity}
  \varphi_{W}({\bf x}+{\bf m}L,x^4)e^{i \pi {\bf n}_{\rm total} \cdot {\bf m}} = \varphi_{W}({\bf x},x^4)\quad ({\bf n}_{\rm total},{\bf m} \in {\bf Z}^3),
\end{equation}
which can be derived from eq.(\ref{eq:factorized_NBS}), together with the periodicity of coordinates ${\bf x_i}\ (i=1,2)$.
Calculations of derivatives (e.g. $\nabla_{\parallel}$ at ${\bf x}_{\parallel} = 0$) are implemented by taking this periodicity into account.
In summary, we can extract the effective LO potential  from  $F_{\phi\phi}(x_1, x_2)$ at sufficiently large $X^4$ through eq.~(\ref{eq:LabHAL_LOpot})
in lattice simulations.

\subsection{Time-dependent method in the laboratory frame} \label{subsect:timedepHAL}
%In lattice QCD simulations, correlation functions except those for pions become noisier at larger $X^4$ in general, so that eq.~(\ref{eq:latNBS_exp}) may not be achieved within small statistical errors.
Since correlation functions in general become noisier at larger $X^4$ in lattice QCD simulations,
it is difficult to achieve eq.~(\ref{eq:latNBS_exp}) within good precision.
To overcome this difficulty,
we have introduced the time-dependent method for an extraction of potentials in the CM frame\cite{HALQCD:2012aa}, which does not require a single state dominance in $F_{\phi\phi}$ such as eq.~(\ref{eq:latNBS_exp}).
%It enables us to study hadron systems suffered from a bad signal-to-noise ratio such as two-baryon system
This method enables us to extract interaction potentials at smaller $X^4$ with less  statistical fluctuations.
We here discuss how we extend the time dependent method to the Lab frame.

A key quantity in the time-dependent method is a normalized correlation function $R$ (we call it ''R-correlator``), which is defined in the Lab frame as
\begin{equation}\label{eq:Rcorr_lab}
  R({\bf x},x^4,X^4) = \frac{F_{\phi\phi, {\bf P}}({\bf x},x^4,X^4)}{F_{\phi,1}(X^4)F_{\phi,2}(X^4)},
\end{equation}
where
\begin{eqnarray}
  F_{\phi\phi, {\bf P}}({\bf x},x^4,X^4) &=& \sum_{\bf X} e^{-i {\bf P \cdot X}}\langle {\rm T} \phi(X+x/2) \phi(X-x/2) {\mathcal J}_{\phi \phi}({\bf P},0) \rangle,
  \label{eq:F_sum} \\
  F_{\phi,i}(X^4) &=& \sum_{\bf x,y} e^{i {\bf p_i \cdot (y-x)}} \langle \phi({\bf x},X^4) \phi^{\dagger}({\bf y},0) \rangle.
\end{eqnarray}
A summation over the CM coordinate ${\bf X}$ with a factor $e^{-i {\bf P \cdot X}}$
in eq.~(\ref{eq:F_sum})
removes the unnecessary plane wave factor $e^{i {\bf P \cdot X}}$ in eq.~(\ref{eq:latNBS}) and reduces statistical fluctuations.
In general, a normalization of the R-correlator, namely the choice of two-point functions in the denominator in eq.~(\ref{eq:Rcorr_lab}), is not unique.
One natural choice is a normalization using the lowest energy in a non-interacting system. % since hadrons in the laboratory frame basically have momenta
%In the laboratory frame hadrons basically have momenta, therefore a normalization by a free lowest energy is one natural choice.
For example, if a source operator is given by $\overline {\mathcal J}_{\phi \phi}({\bf P},0) = \sum_{\bf x,y} e^{i {\bf p_1 \cdot x}} e^{i {\bf p_2 \cdot y}} \bar \phi({\bf x},0) \bar \phi({\bf y},0)$
with ${\bf P} = {\bf p_1 + p_2}$,
%then the normalization is done by $C_1(X^4) = \langle \varphi_1({\bf P},X^4) \bar \varphi_1({\bf P},0) \rangle $ and $C_2(X^4) = \langle \varphi_2({\bf 0},X^4) \bar \varphi_2({\bf 0},0) \rangle $.
we then take $F_{\phi,1}(T) = \sum_{\bf x,y} e^{i {\bf p_1 \cdot (y-x)}} \langle \phi({\bf x},T) \bar \phi({\bf y},0) \rangle$
and $F_{\phi,2}(T) = \sum_{\bf x,y} e^{i {\bf p_2 \cdot (y-x)}} \langle \phi({\bf x},T) \bar \phi({\bf y},0) \rangle$ in eq.~\eqref{eq:Rcorr_lab}.
Alternatively, we may choose hadron masses measured in  the center-of-mass frame for the normalization.
While such a difference in normalizations does not affect final results of  potentials in the continuum limit,
it may produce some systematic uncertainties in potentials due to discretization errors at finite lattice spacings.
Since meson energies in the laboratory frame suffer from larger discretization effects due to non-zero momenta,
it is important to keep such effects under control in our numerical simulation, which will be investigated in Appendix~\ref{appx:lab_hal_sys}.
In the rest of our paper, we employ the normalization using the lowest energy in the non-interacting system.

To extract the potential from the R-correlator, we first define
\begin{eqnarray}
  G({\bf x},x^4,X^4) &=& \left( (\partial_{X^4} - W_{0,{\rm free}})^2 - {\bf P}^2 \right) R({\bf x},x^4,X^4),
  \label{eq:def_G}\\
  E({\bf x},x^4,X^4) &=& \frac{1}{4m} \left[ \partial_{X^4}^2 - 2 W_{0,{\rm free}} \partial_{X^4} + W_{0,{\rm free}}^2 - {\bf P}^2 - 4m^2 \right] G({\bf x},x^4,X^4), \label{eq:def_E}\\
  L_{\perp}({\bf x},x^4,X^4) &=& \nabla_{\perp}^2 G({\bf x},x^4,X^4) ,\label{eq:lapB}\\
  L_{\parallel}({\bf x},x^4,X^4) &=& \left( - (\partial_{X^4} - W_{0,{\rm free}}) \nabla_{\parallel} + i {\bf P} \partial_{x^4} \right)^2 R({\bf x},x^4,X^4), \label{eq:lapC}
\end{eqnarray}
where $W_{0,{\rm free}}$ is a non-interacting energy level used in the normalization.
At a moderately large $X^4$ where inelastic contributions can be neglected,
we have
\begin{eqnarray}
 G({\bf x},x^4,X^4) &\simeq& \sum_{\bf n} B'_{\bf n} W_{{\rm CM},{\bf n}}^2 \varphi_{W_{\bf n}}({\bf x},x^4) e^{- (W_{\bf n} - W_{0,{\rm free}}) X^4}, \\
  E({\bf x},x^4,X^4) &\simeq&  \sum_{\bf n} B'_{\bf n} W_{{\rm CM},{\bf n}}^2 \frac{k_{n}^{*2}}{m} \varphi_{W_{\bf n}}({\bf x},x^4) e^{- (W_{\bf n} - W_{0,{\rm free}}) X^4},
  \label{eq:k2n} \\
  L_{\perp}({\bf x},x^4,X^4) &\simeq&  \sum_{\bf n} B'_{\bf n} W_{{\rm CM},{\bf n}}^2 \nabla_{\perp}^2 \varphi_{W_{\bf n}}({\bf x},x^4) e^{- (W_{\bf n} - W_{0,{\rm free}}) X^4}, \\
  L_{\parallel}({\bf x},x^4,X^4) &\simeq& \sum_{\bf n} B'_{\bf n} W_{{\rm CM},{\bf n}}^2 \gamma_{\bf n}^2 (\nabla_{\parallel} + i {\bf v}_{\bf n} \partial_{x^4})^2 \varphi_{W_{\bf n}}({\bf x},x^4) e^{- (W_{\bf n} - W_{0,{\rm free}}) X^4},
  \label{eq:gv2n}
\end{eqnarray}
where
\begin{eqnarray}
B'_{\bf n} &=& {B_{\bf n}\over {C_1 C_2}}, \quad F_{\phi,i}(X^4)= C_i e^{-\sqrt{m^2+p_i^2} X^4} +\mbox{(inelastic contributions)}, \\
W_{{\rm CM},{\bf n}}^2 &=& W^2_{\bf n} - {\bf P}^2 = 4(k_{\bf n}^{*2}+m^2), \quad
\gamma_{\bf n}^2 = {W_{\bf n}^2 \over W_{{\rm CM},{\bf n}}^2}.
\end{eqnarray}

By combining these  and eq.(\ref{eq:LabHAL_pot}), we obtain
\begin{eqnarray}
  \left. \left(\frac{L_{\perp} + L_{\parallel}}{m} + E \right)({\bf x},x^4,X^4) \right|_{x^4 = 0, {\bf x}_{\parallel}=0} &\simeq&  \sum_i V^{i}_{x^{*4}=0} \left({\bf x}^*_{\perp}={\bf x}_{\perp},{\bf x}^*_{\parallel}=0 \right) \nonumber  \\
&\times& \left.  \left( (\nabla^{*2})^i G(x, x^4, X^4) \right) \right|_{x^4 = 0, {\bf x}_{\parallel}=0}
\label{eq:LabHAL_tdeppot}
\end{eqnarray}
at a moderately large $X^4$,
where an operation of starred-Laplacians on $G$ is understood as
\begin{equation}
  (\nabla^{*2})^i G(x, x^4, X^4) = \sum_{\bf n} B'_{\bf n} W_{{\rm CM},{\bf n}}^2 \left(\nabla_{\perp}^2 + \gamma_{\bf n}^2 (\nabla_{\parallel} + i {\bf v}_{\bf n} \partial_{x^4})^2 \right)^i \varphi_{W_{\bf n}}({\bf x},x^4) e^{- (W_{\bf n} - W_{0,{\rm free}}) X^4}.
\end{equation}
Note that we can move $V^i$ outside a summation over ${\bf n}$ for elastic states in eq.(\ref{eq:LabHAL_tdeppot}) only at $x^{*4} = 0$,
since a scheme of the potential depends on ${\bf n}$ through $\gamma_{\bf n}$ unless $x^{*4} =
\gamma_{\bf n}(x^4- i{\bf v}_{\bf n} \cdot {\bf x}_{\parallel})=0$.
This procedure is more complicated than
the conventional time-dependent method~\cite{HALQCD:2012aa},
since we need to sum over ${\bf n}$
without knowing not only $k_{\bf n}^{*2}$  in eq.~(\ref{eq:k2n}) but also Lorentz factor $\gamma_{\bf n}^2$ and velocity ${\bf v}_{\bf n}$ in eq.~(\ref{eq:gv2n}) for each ${\bf n}$, by combining several terms as shown above.

Finally, the effective LO potential in the time-dependent method reads
\begin{equation} \label{eq:effLOpot}
  V^{\rm LO}_{x^{*4}=0}({\bf x}_{\perp}) = \left. \frac{\left(L_{\perp} + L_{\parallel} + mE \right)({\bf x},x^4,X^4)}{mG({\bf x},x^4,X^4)} \right|_{x^4=0, {\bf x}_{\parallel}=0}.
\end{equation}

%%%%% method end %%%%%

%%%%% simulation details and result %%%%%
\section{Numerical results in the $I=2$ $\pi \pi$ system} \label{sect:demonstration}
%In this section, we demonstrate the HAL QCD method with non-zero total momenta in the simple $I=2$ $\pi \pi$ system. Firstly, we give some details of our calculation, then we show the numerical results. We compare our results with the conventional HAL QCD results in CM frame.

\subsection{Calculation of correlation functions} \label{subsect:opsetting}
Let us consider the $I=2$ $\pi \pi$ S-wave scattering as an explicit example.
We define correlation functions of this system as
\begin{eqnarray} \label{eq:4ptfunc}
  F_{\pi^+ \pi^+, {\bf P}}({\bf x},x^4,X^4) &=& \sum_{\bf X} e^{-i {\bf P \cdot X}}\langle {\rm T} \pi^{+}(X+x/2) \pi^{+}(X-x/2) {\mathcal J}_{\pi^+ \pi^+}({\bf P},0) \rangle , \\
  F_{\pi^+,i}(X^4) &=& \sum_{\bf x,y} e^{i {\bf p_i \cdot (y-x)}} \langle \pi^+({\bf x},X_4) \pi^-({\bf y},0) \rangle,
\end{eqnarray}
where the positively (negatively) charged pion operator is constructed in terms of up and down quark fields $u(x)$ and $d(x)$ as
$\pi^{+}(x) = \bar d(x) \gamma_5 u(x)$ ($\pi^-(x) = \bar u(x) \gamma_5 d(x)$).
%Total momenta are chosen as ${\bf P} = (0,0,2 \pi / L \times n)\ (n = 0, 1, 2)$, and the corresponding source operators in our calculation are given as
%{Total momenta are chosen as ${\bf P} = \frac{2 \pi}{L}{\bf n}_{\rm total} = \frac{2 \pi}{L} (0,0,n)\ (n = 0, 1, 2)$, and the corresponding source operators in our calculation are given as}
{Total momenta are chosen as ${\bf P} = (P_x, P_y, P_z) = \frac{2 \pi}{L} (0,0,n)\ (n = 0, 1, 2)$, and corresponding source operators are given by}
\if0
\begin{eqnarray}
  {\mathcal J}_{\pi^+ \pi^+} \left({\bf P} = 0 ,0 \right) &=& \overline \pi_s^+ \left({\bf p_1} = 0,0\right) \overline \pi_s^+ \left({\bf p_2} = 0,0 \right), \\
  {\mathcal J}_{\pi^+ \pi^+} \left({\bf P} = \frac{2 \pi}{L}{\bf e_z} ,0 \right) &=& \overline \pi_s^+ \left({\bf p_1} = \frac{2 \pi}{L}{\bf e_z},0\right) \overline \pi_s^+ \left({\bf p_2} = 0,0 \right), \\
  {\mathcal J}_{\pi^+ \pi^+} \left({\bf P} = \frac{2 \pi}{L}{\bf e_z} \times 2 ,0 \right) &=& \overline \pi_s^+ \left({\bf p_1} = \frac{2 \pi}{L}{\bf e_z},0 \right) \overline \pi_s^+ \left({\bf p_2} = \frac{2 \pi}{L}{\bf e_z},0 \right),
\end{eqnarray}
\fi
{
\begin{eqnarray}
  {\mathcal J}_{\pi^+ \pi^+} \left({\bf P} = (0,0,0) ,0 \right) &=& \overline \pi_s^+ \left({\bf p_1} = {\bf 0},0\right) \overline \pi_s^+ \left({\bf p_2} = {\bf 0},0 \right) \\
  {\mathcal J}_{\pi^+ \pi^+} \left({\bf P} = (0,0,1) ,0 \right) &=& \overline \pi_s^+ \left({\bf p_1} = (0,0,1),0\right) \overline \pi_s^+ \left({\bf p_2} = {\bf 0},0 \right) \\
  {\mathcal J}_{\pi^+ \pi^+} \left({\bf P} = (0,0,2) ,0 \right) &=& \overline \pi_s^+ \left({\bf p_1} = (0,0,1),0 \right) \overline \pi_s^+ \left({\bf p_2} = (0,0,1),0 \right),
\end{eqnarray}
}
where all momenta are given in unit of $2 \pi / L$ and we keep this notation in the remaining of this paper.
A pion creation operator with momentum is defined as
\begin{equation}
  \overline \pi^+({\bf p},0) = \sum_{\bf y} \pi^-({\bf y},0) e^{+i {\bf p\cdot y} },
\end{equation}
{
%Subscripts $s$ in the pion operators indicate that they consist of smeared quarks to suppress the inelastic contributions at earlier imaginary time. Details of the smearing are given in Sect.~\ref{subsect:simudetail}.
and a subscript $s$ indicates that quark fields in operators are smeared to suppress inelastic contributions at earlier imaginary time, as will be explained in Sect.~\ref{subsect:simudetail}.
}
{
For calculations of correlation functions defined above, we employ the one-end trick\cite{McNeile:2006bz}, which enables us to evaluate a combination of two all-to-all propagators with spatial summation by a single stochastic estimator.
}
%{In the following, each results are labeled by the total momentum in unit of $2 \pi / L$, as $P(0,0,n)\ (n = 0, 1, 2)$.}
%In general, we need all-to-all propagators to calculate correlation functions involving source operators with momenta.
%To handle this, we employ the one-end trick\cite{}, which enables us to calculate the combination of two all-to-all propagators with spatial summation by a single stochastic estimator.
%We give a brief review on the one-end trick and details of calculation of $F_{\pi^+ \pi^+, {\bf P}}$ in Appendix~\ref{appex:oneend} and \ref{appex:corrfunc}, respectively.

\subsection{Simulation details} \label{subsect:simudetail}
We employ 2+1 flavor full QCD configurations generated by CP-PACS Collaborations~\cite{Aoki:2008sm} on a $32^3 \times 64$ lattice with the Iwasaki gauge action~\cite{Iwasaki:1985we} at $\beta$ = 1.90 and
the non-perturbatively improved Wilson-clover action~\cite{Sheikholeslami:1985ij} at $c_{\rm SW}$ = 1.7150,
which corresponds to  the lattice spacing $a = 0.0907$ fm.
Hopping parameters $(\kappa_{ud},\kappa_s) = (0.13700,0.13640)$ lead to the pion mass $m_{\pi} \approx 700$ MeV.
A periodic boundary condition is employed in all spacetime directions.
Correlation functions with ${\bf P} \neq 0$ (${\bf P} = 0$) are evaluated by 399 gauge configurations $\times$ 64 timeslices (399 gauge configurations $\times$ 16 timeslices), and we estimate statistical errors by the jack-knife method with a bin-size 21 in all cases.
We label results with $P=(0,0,n) \ (n = 0, 1, 2)$ as CM, case 1, and case 2, respectively.
Hereafter dimensionful quantities without corresponding unit are written in lattice unit unless otherwise stated.
%Table.~\ref{} and \ref{} show the general setups of our study and details of the one-end trick, respectively.

We employ the smeared quark source $q_s({\bf x},t) = \sum_{\bf y}f({\bf x} - {\bf y})q({\bf y},t)$ with the Coulomb gauge fixing,
%to suppress the inelastic contributions at earlier imaginary time.
where the smearing function $f({\bf x})$ is taken as~\cite{Iritani:2016jie}
\begin{equation}
  f({\bf x}) = \begin{cases}
    Ae^{-B|{\bf x}|} & (0 < |{\bf x}| < (L-1)/2)\\
    1 & (|{\bf x}| = 0)\\
    0 & (|{\bf x}| \geq (L-1)/2)
  \end{cases}
\end{equation}
with $A = 1.2$, $B = 0.30$.
We generate a single $Z_4$ noise vector for each insertion of the one-end trick.
To reduce stochastic noise contaminations from noise vectors, the dilution technique~\cite{Foley:2005ac} is applied to color, spinor and spatial indices. We fully dilute color and spinor indices, and $s2$ (even-odd) dilution~\cite{Akahoshi:2020ojo} is taken for the spatial index.
Numerical derivatives are approximated by 2nd order differences.
%We prepare three correlation functions with $x^4 = 0, \pm 2$ to estimate $x^4$ derivatives, and
In estimations of $x^4$ derivatives, we utilize correlation functions with even relative time $x^4 = 0, \pm 2$ to keep the CM time $X^4$ integer.

\begin{figure}[htbp]
  \begin{tabular}{cc}
    \begin{minipage}{0.5\hsize}
        \centering
        \includegraphics[width=0.9\hsize,clip]{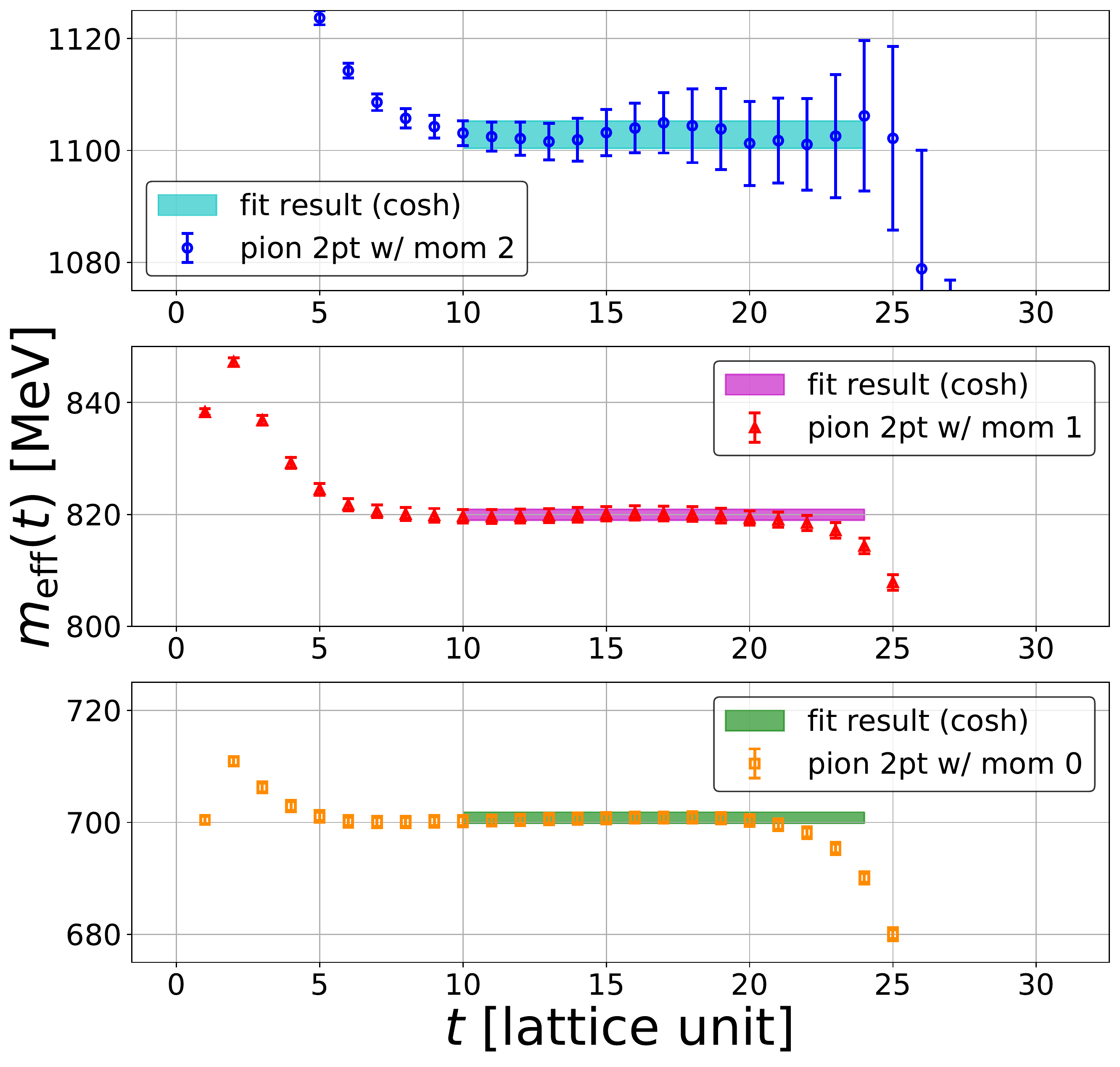}
    \end{minipage}
    &
    \begin{minipage}{0.5\hsize}
        \centering
        \includegraphics[width=\hsize,clip]{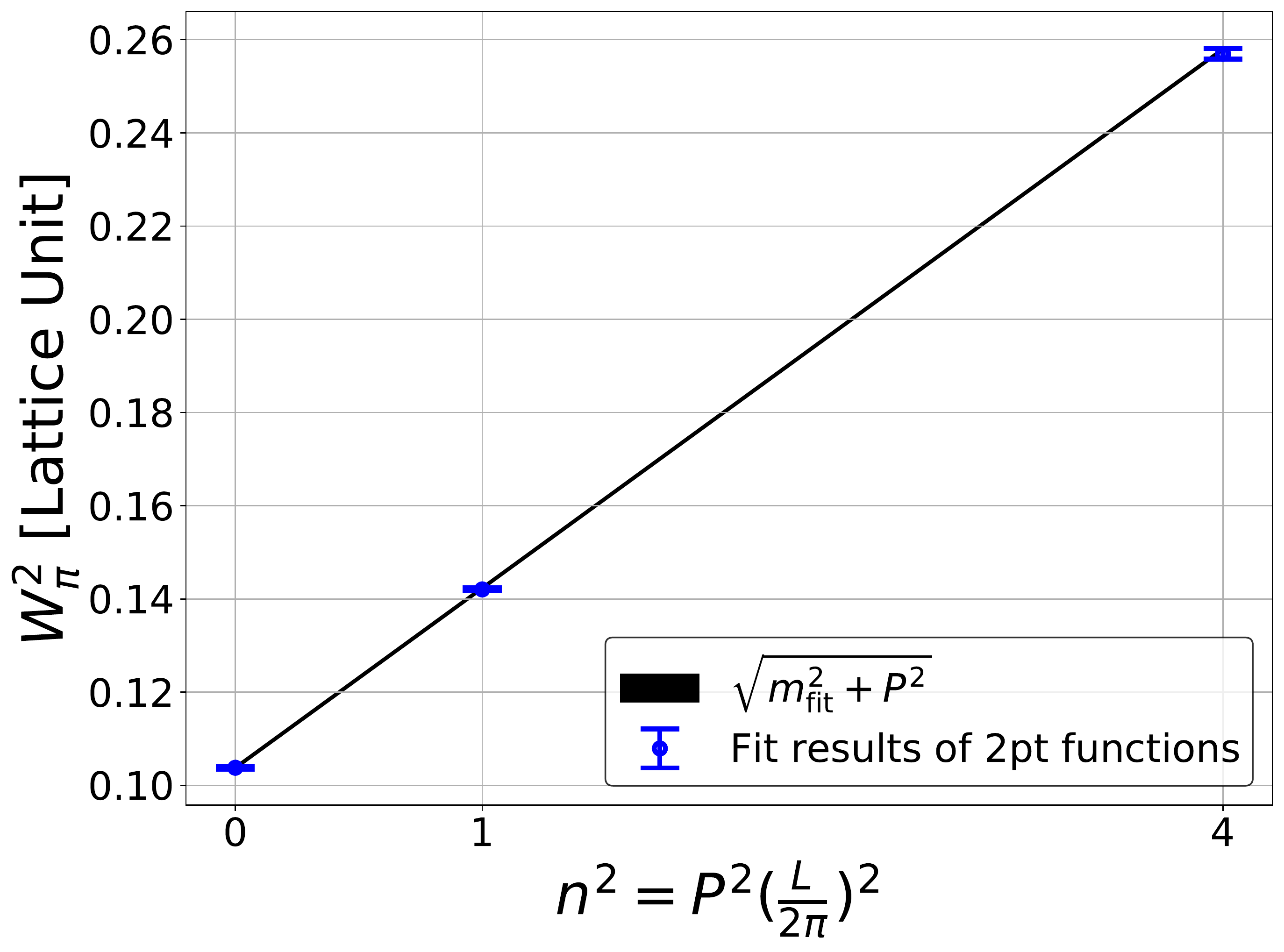}
    \end{minipage}
  \end{tabular}
  \caption{
  (Left) Effective energies of a single pion with $|p| = 0$ (orange square), $|p| = 2 \pi / L$ (red triangle) and $|p| = 4 \pi / L$ (blue circle). fit results using a cosh function and corresponding fit ranges are shown by light color bands.
  (Right) A comparison between extracted energy levels (blue points) and the continuum dispersion relation, $E(p) = \sqrt{m_{\pi,{\rm fit}}^2 + p^2}$ (black solid line). For $m_{\pi,{\rm fit}}$, we use the fit result of 2pt correlation function with $|p| = 0$ (green band in the bottom of the left figure).
  }
  \label{fig:effenes}
\end{figure}
{
Since our formulation relies on the continuum dispersion relation, we first check a behavior of a pion dispersion relation.
Figure~\ref{fig:effenes} (Left) shows effective energies of a single pion with momenta
%up to $|p| = 4 \pi / L$
{${\bf p} = (0,0,n)\ (n = 0, 1, 2)$}.
We obtain good plateaus for all cases thanks to the quark smearing.
%{The pion mass in the continuum dispersion relation is determined by the cosh-shape fit of pion 2pt correlator with $|p| = 0$.}
An energy eigenvalue for each momentum channel is extracted by a single cosh fit, as shown by light color bands in Fig.~\ref{fig:effenes} (Left).
We then compare those to the continuum dispersion relation, $E(p) = \sqrt{m_{\pi}^2 + {\bf p}^2}$.
As seen in Fig.~\ref{fig:effenes} (Right), extracted energy levels (blue points) are consistent with the continuum dispersion relation (black solid line) up to $|p| = 4 \pi / L$, so that we can safely utilize the continuum dispersion relation in this study.
}

\subsection{The NBS wave function in the laboratory frame} \label{subsect:NBSwave}
\begin{figure}[htbp]
  \begin{tabular}{cc}
    \begin{minipage}{0.5\hsize}
        \centering
        \includegraphics[width=\hsize,clip]{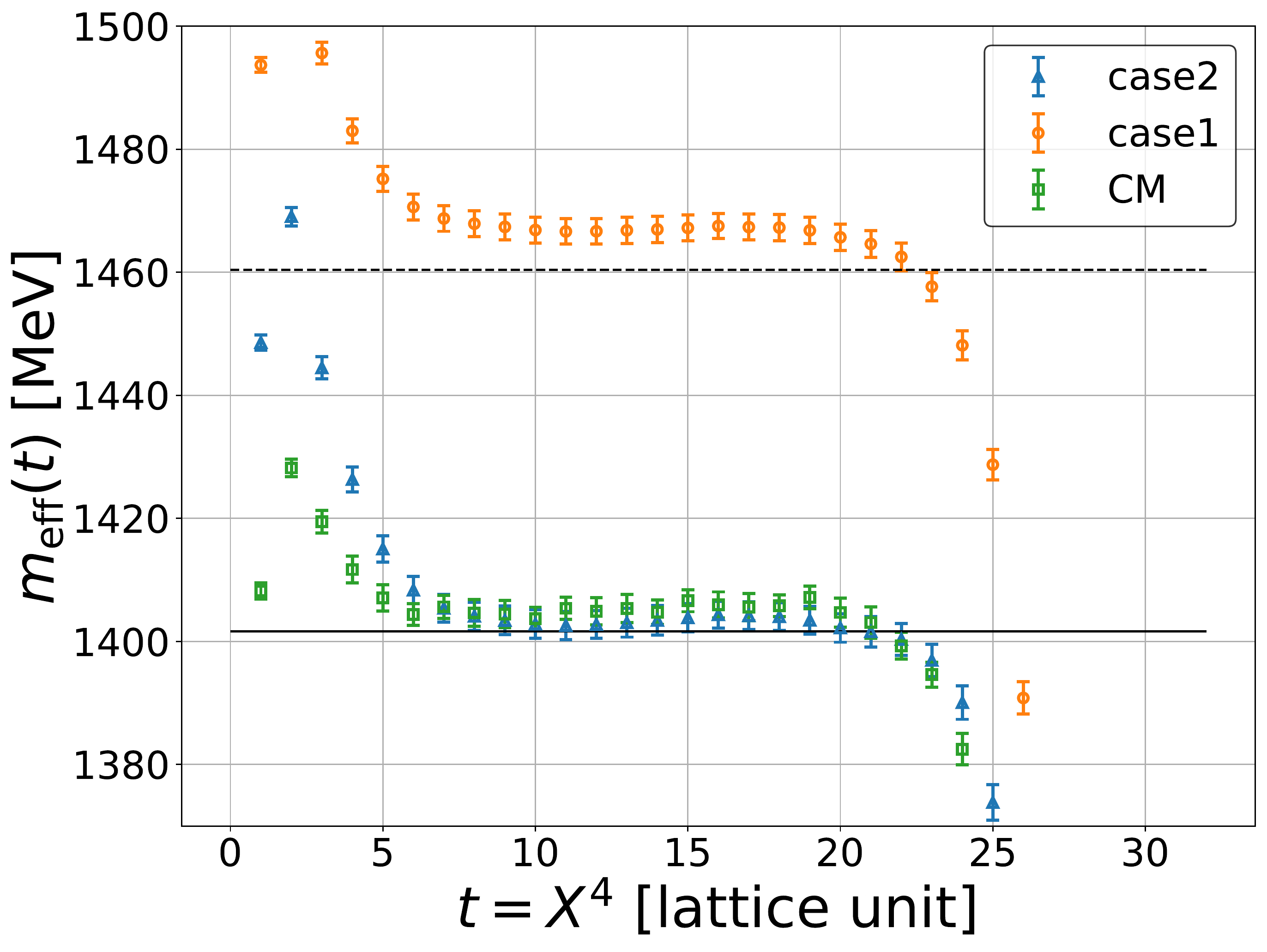}
    \end{minipage}
    &
    \begin{minipage}{0.5\hsize}
        \centering
        \includegraphics[width=\hsize,clip]{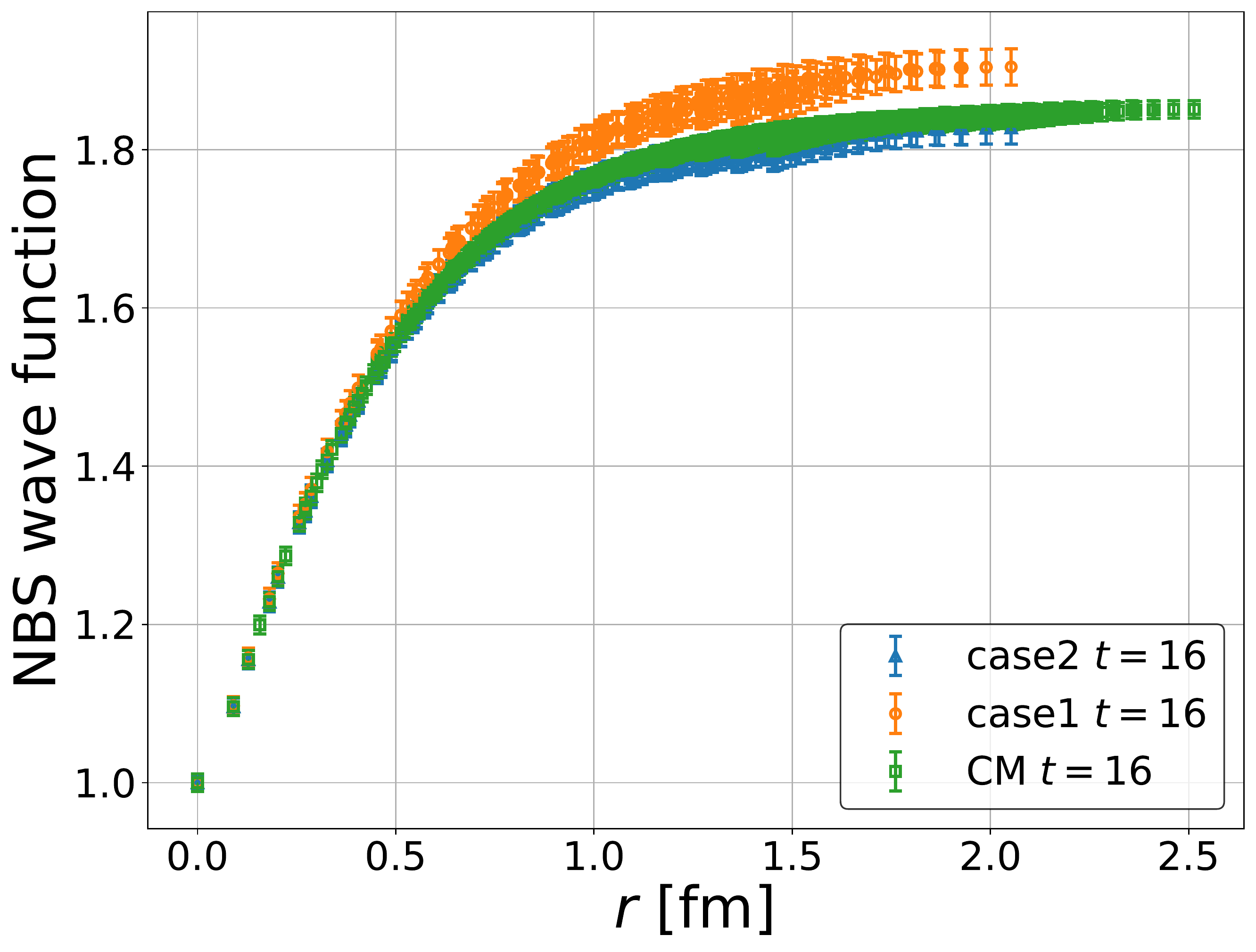}
    \end{minipage}
  \end{tabular}
  \caption{(Left) Effective energies obtained by spatial summation of $F_{\pi^+ \pi^+, {\bf P}}$.
  Laboratory frame energies are boosted back to the CM frame via $W_{\rm CM}^2 = W_{\rm L}^2 - P^2$.
  Dashed and solid lines show the lowest non-interacting energies with ${\bf P} = (0,0,2\pi/L)$ and $(0,0,4\pi/L)$, respectively.
  (Right) spatial dependence of $F_{\pi^+ \pi^+, {\bf P}}$. For $|P| \neq 0$, we fix ${\bf x}_{\parallel}={\bf 0}$ and only show ${\bf x}_{\perp}$ dependence. Values at the origin are normalized to unity. }
  \label{fig:i2pp_NBS_comp}
\end{figure}
{
Before discussing the interaction potential, let us consider the behavior of $F_{\pi^+ \pi^+, {\bf P}}$ defined in eq.(\ref{eq:4ptfunc}).
{As already discussed in eq.~(\ref{eq:latNBS_exp0}) and (\ref{eq:latNBS_exp}), we can extract ground state energies and corresponding NBS wave functions at $X^4 \gg 1$.}
Figure~\ref{fig:i2pp_NBS_comp} (Left) shows effective energies obtained from the $X^4$ dependence of $\sum_{\bf x} F_{\pi^+ \pi^+, {\bf P}}({\bf x},x^4=0,X^4)$ {and corresponding non-interacting energy levels (horizontal lines).}
We observe that all effective energies reach plateaus {at around $X^4 = 15$}, and they slightly shift upward from the non-interacting energy levels.
{It indicates that
$F_{\pi^+ \pi^+, {\bf P}}$ is almost dominated by the ground state at that timeslice and
the interaction is repulsive as reported in previous studies~\cite{Sasaki:2013vxa,Kawai:2017goq,Akahoshi:2019klc}.}
%In the following, we show the results at $X^4 = 15$ unless otherwise stated.
\if0
The repulsive behaviors are also confirmed by ${\bf x}_{\perp}$ dependence of $F_{\pi^+ \pi^+, {\bf P}}$ with $({\bf x}_{\parallel},x^4)=({\bf 0},0)$ shown in Fig.~\ref{fig:i2pp_NBS_comp} (Right), which is directly related to the NBS wave function in the CM frame as
\begin{equation}
  \begin{split}
    F_{\pi^+ \pi^+, {\bf P}}({\bf x}_{\perp},{\bf x}_{\parallel}=0,x^4=0,X^4) &\approx \sum_n B_n \varphi_{W_n}({\bf x}_{\perp},{\bf x}_{\parallel}=0,x^4=0) e^{-W_n X^4} \\
    & = \sum_n B_n \varphi_{W^*_n}({\bf x}^*_{\perp}={\bf x}_{\perp},{\bf x}^*_{\parallel}=0,x^{*4}=0) e^{-W_n X^4},
  \end{split}
\end{equation}
here we omit the inelastic contributions.
We observe that the NBS wave function in case 2 is more similar to that in the CM, % than that in case 1,
which is expected by the fact that the lowest energy in case 2 is almost the same as that in the CM frame.
\fi
{
In the Figure~\ref{fig:i2pp_NBS_comp} (right), we show the spatial dependence of the $F_{\pi^+ \pi^+, {\bf P}}({\bf x}_{\perp},{\bf x}_{\parallel}=0,x^4=0,X^4)$ at $t = X^4 = 16$,
which is expected to approach the ground state NBS wave function in the CM frame in accordance with eq.(\ref{eq:latNBS_exp}) as
\begin{equation}
  \begin{split}
    F_{\pi^+ \pi^+, {\bf P}}({\bf x}_{\perp},{\bf x}_{\parallel}=0,x^4=0,X^4) &\approx B_{\rm min} \varphi_{W_{\rm min}}({\bf x}_{\perp},{\bf x}_{\parallel}=0,x^4=0) e^{-W_{\rm min} X^4} \\
    & = B_{\rm min} \varphi_{W^*_{\rm min}}({\bf x}^*_{\perp}={\bf x}_{\perp},{\bf x}^*_{\parallel}=0,x^{*4}=0) e^{-W_{\rm min} X^4}.
    \end{split}
\end{equation}
A small number of data points in $|P| \neq 0$ is due to the condition of the equal-time scheme, $({\bf x}_{\parallel},x^4)=({\bf 0},0)$.
As expected by the behavior of effective energies, they show the monotonic increasing behaviors in $r$, typical for the repulsive force.
%behavior by the monotonic increase of the NBS wave function.
Moreover,  the NBS wave function with $|P| = 4\pi / L$ (case 2) is very similar to that with $|P| = 0$ (CM),
probably due to a fact that the lowest energy with $|P| = 4\pi / L$ boosted back to the CM frame is roughly equal to the one with $|P| = 0$, as seen in Fig.~\ref{fig:i2pp_NBS_comp} (Left).
}
}

\subsection{Effective leading-order potential} \label{subsect:numepot}
\begin{figure}[htbp]
  \centering
  \includegraphics[width=100mm,clip]{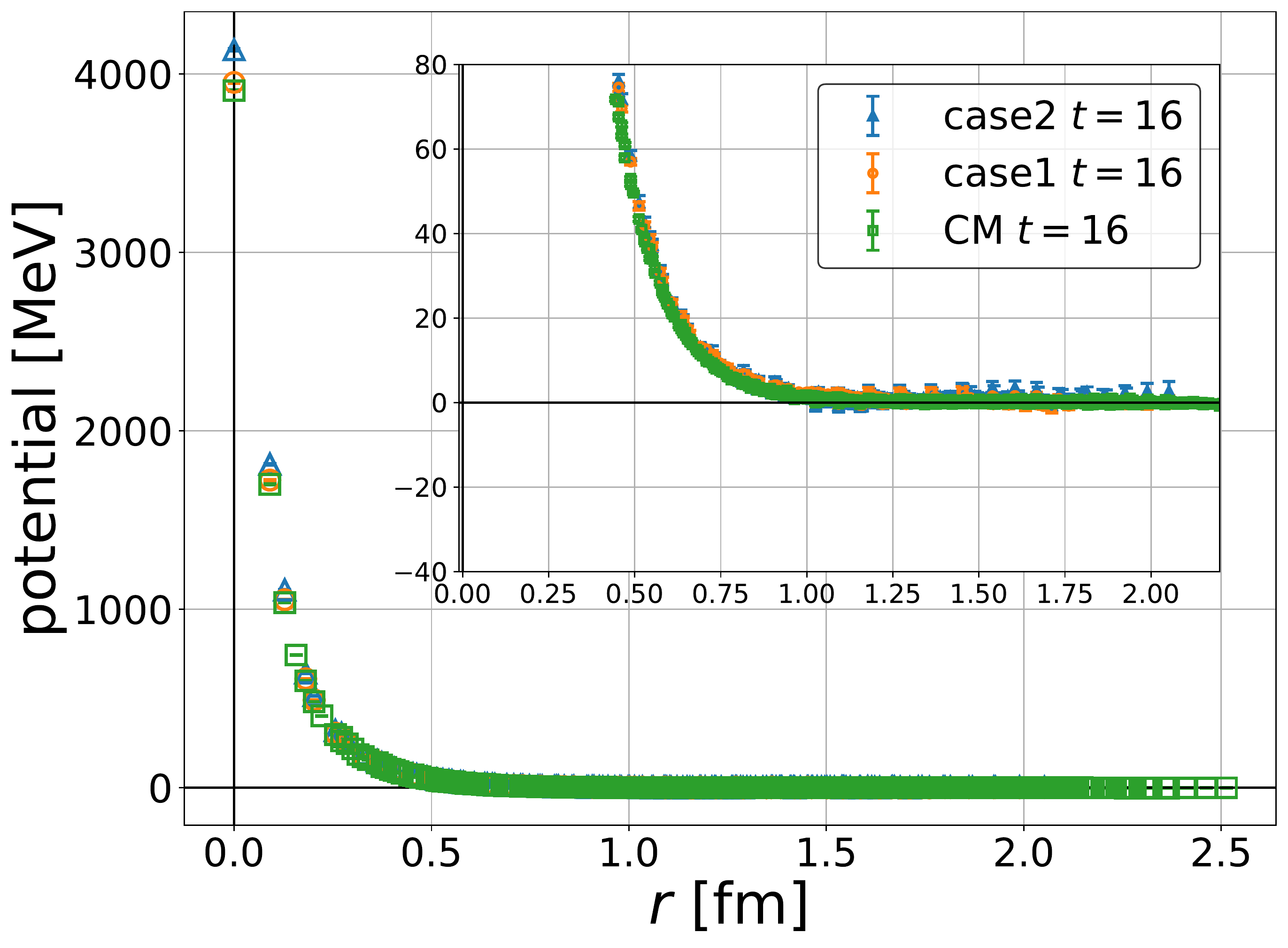}
  \caption{A comparison of all effective LO potentials. Inset shows an enlarged view of the potentials.}
  \label{fig:i2pp_pot_compall}
\end{figure}
\begin{figure}[htbp]
  \begin{tabular}{cc}
    \begin{minipage}{0.5\hsize}
        \centering
        \includegraphics[width=\hsize,clip]{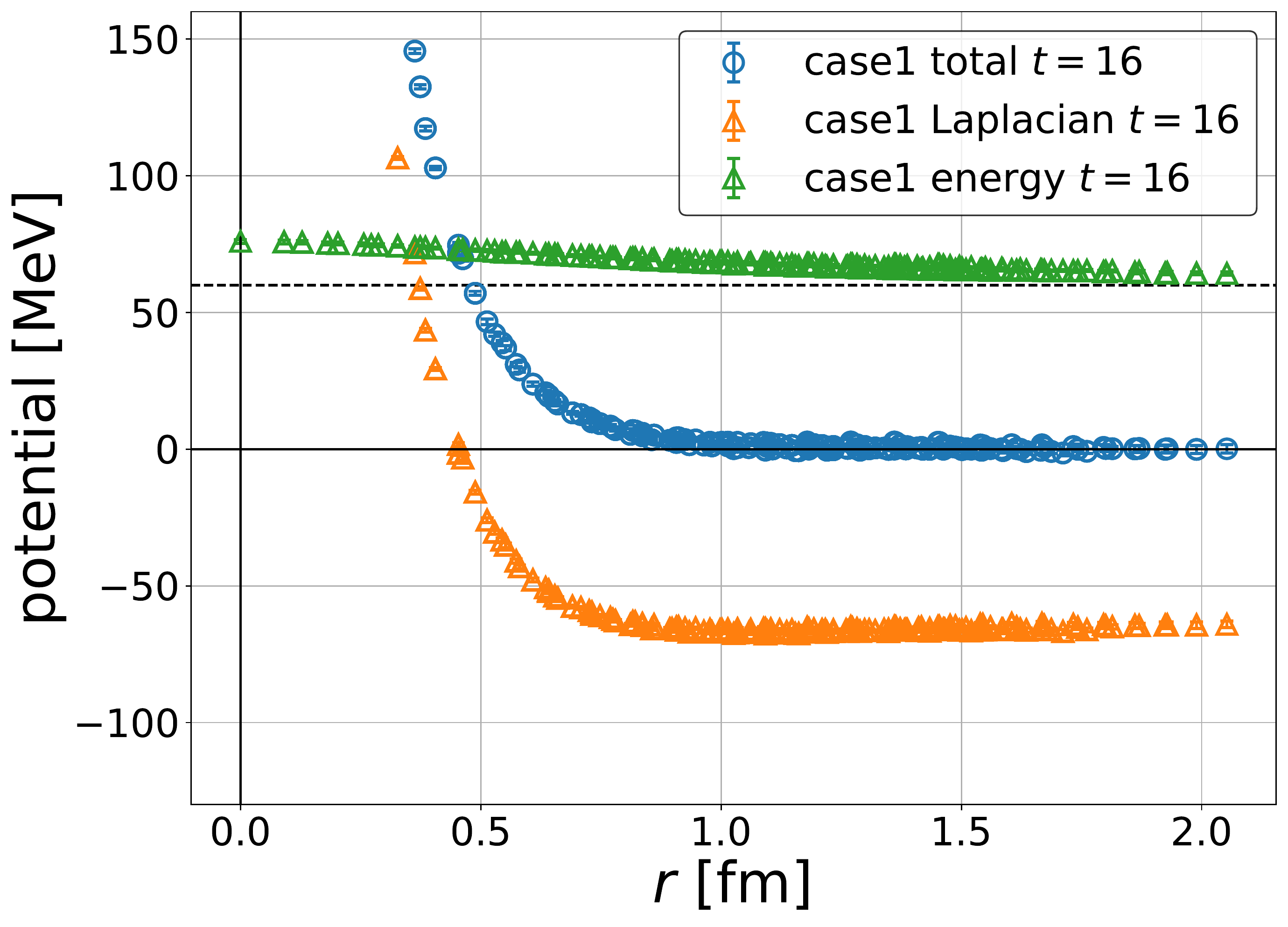}
    \end{minipage}
    &
    \begin{minipage}{0.5\hsize}
        \centering
        \includegraphics[width=\hsize,clip]{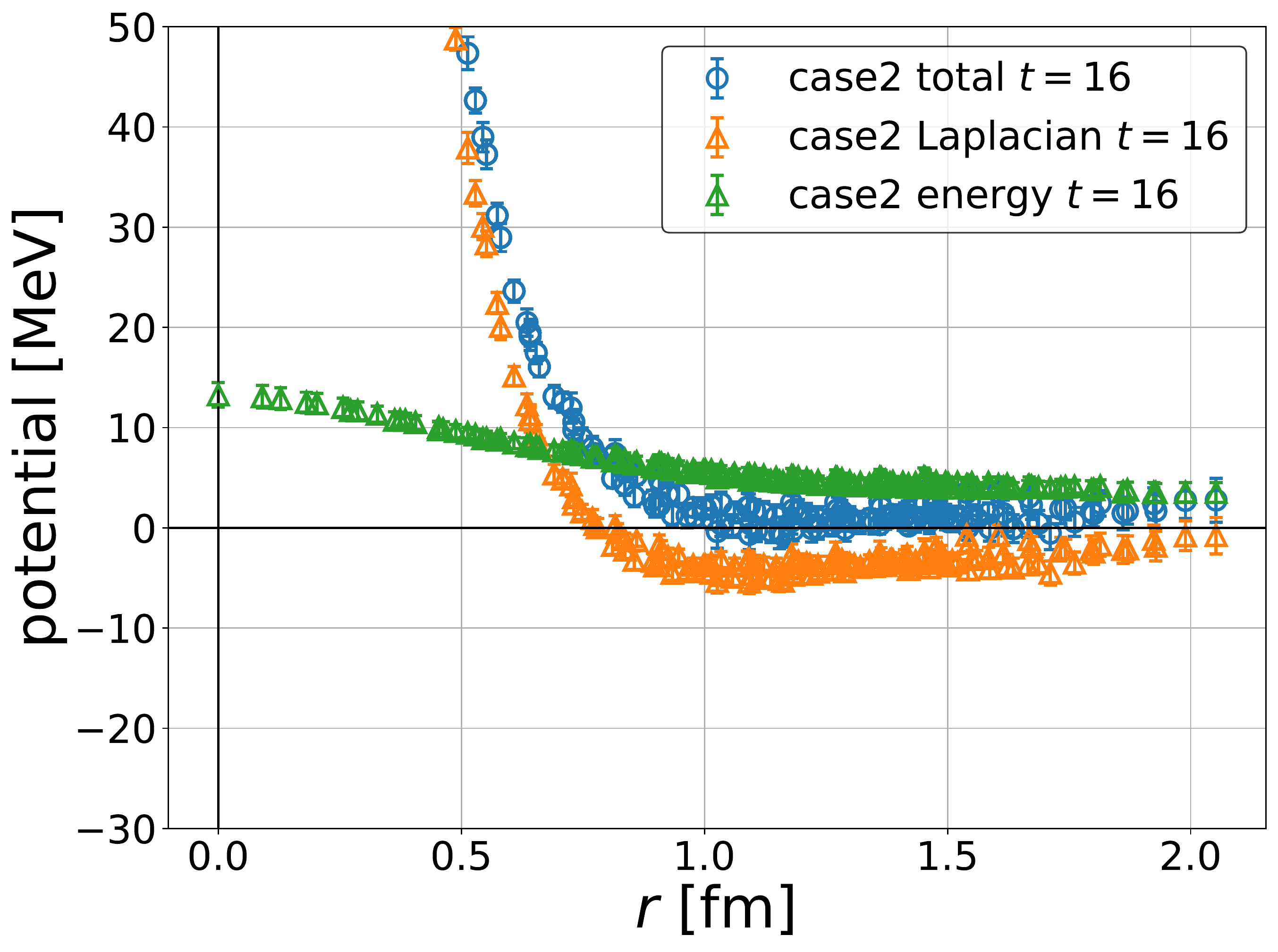}
    \end{minipage}
  \end{tabular}
  \caption{A decomposition of the effective LO potential in case 1 (Left) and case 2 (Right).
  A dotted line represents an expected relative energy in a non-interacting case.}
  \label{fig:i2pp_pot_term}
\end{figure}
{
{We next consider effective leading order (LO) potentials obtained by the time-dependent method.}
%{In the following, we show results at $t = X^4 = 16$ unless otherwise stated, since we can expect that the inelastic contributions are well suppressed and the ground states dominate the correlation functions at that timeslice.}
{
Figure~\ref{fig:i2pp_pot_compall} shows effective LO potentials with three different total momenta at $t = X^4 = 16$.
}
As already discussed in Sect.~\ref{subsect:NBSwave}, potentials show repulsive behaviors, and they are consistent with each other except at short distances.
A small difference observed at short distances may be explained by finite lattice spacing effects.
%, since the laboratory frame calculation has more numerical derivatives than the center-of-mass calculation.

We also observe that potentials in cases 1 and 2 have larger statistical errors
and non-smooth behaviors as compared with that in the CM case.
Typically, introducing non-zero momentum makes correlation functions noisier, since an enhancement of statics by the translational invariance is reduced.
Indeed, we have already observed that NBS wave functions themselves are noisier than the one in the CM frame (see Fig.\ref{fig:i2pp_NBS_comp} (right)).
In addition, larger statistical fluctuations  in the laboratory frame are probably caused also by 4th-order $X^4$ derivative terms in the time-dependent method. To estimate 4th-order $X^4$ derivatives at a fixed $X^4$ by the numerical difference, we have to utilize correlation functions at $X^4 \pm 2$, which are absent for 2nd-order derivatives.
%we only need information at $X^4 \pm 1$ for estimation of 2nd-order derivatives).
Since data at larger $X^4$ are generally nosier, 4th-order $X^4$ derivatives are expected to be noisier as well.
%The non-smooth behaviors can be understood by the partial wave contents of the irreducible representation of the cubic group.
Non-smooth behaviors for potentials in the laboratory frame, on the other hand, may be explained by a contamination from the  $l=2$ partial wave, which is absent in the CM frame, as follows.
In the laboratory frame calculation, the cubic rotation is no longer the symmetry of the system, since the cubic box is deformed by
the Lorentz contraction if it is boosted back to the CM frame.
In our setup, since the box becomes a rectangular with size $L \times L \times \gamma L$ with a boost factor $\gamma$ in the CM frame, the cubic symmetry is reduced to the one which makes this rectangular intact.
%rotational symmetry reduces to a subgroup of the cubic symmetry which keeps the rectangle invariant.
An irreducible representation $A_1^+$ of the reduced symmetry contains contributions from angular momenta $l = 0, 2, \cdots$,
in contrast to the $A_1^+$ in the cubic symmetry, which allows $l=0,4,6,\cdots$ partial waves.
Since lower partial waves in general have larger contributions at low energy,
a contamination form the  $l = 2$ partial wave causes non-smooth behaviors of potentials in the laboratory frame,
which is stronger than the one by  the $l=4$ partial wave, the lowest partial wave contamination in  the CM frame.

%but statistical errors of Lab frame calculations become somewhat larger than that of the CM frame calculation.
%{Such enhancements of statistical errors mainly come from the lower statistics of $F_{\pi^+ \pi^+, {\bf P}}$.}
%{Therefore, more statistical samples may be mandatory to obtain precise potentials with non-zero total momenta in more complicated hadronic systems, although it is not the case for $I=2$ $\pi\pi$ system.}
To see how the time-dependent method works in detail, we decompose potentials into the Laplacian term, $\dfrac{L_{\perp} + L_{\parallel}}{mG}$ in eq.~(\ref{eq:effLOpot}), and the energy term, $\dfrac{mE}{mG}$ in eq.~(\ref{eq:effLOpot}), as shown in Fig.~\ref{fig:i2pp_pot_term}.
We observe that the Laplacian term (orange) and the energy term (green) are away from zero, but the total (blue) converges to zero at large distances thanks to their cancellations.
Since values of the energy shift (green) roughy agree with expectations from lowest energies in non-interacting cases, the cancellation of two terms (orange and green) is a strong evidence on the validity of the time-dependent method with non-zero total momenta.
%in Fig.~\ref{fig:i2pp_pot_term},
%From these, we confirm that we can construct interaction potentials from the Lab frame NBS wave functions in practice.%,
%{even though we may need more statistics to suppress enhanced statistical errors in more complicated situations.}
}
%In the following analysis, we use the potentials at $X^4 = 10$, the largest time with small time dependence of potentials, and estimate the systematic errors from time-dependence if possible.
%Although there are still some points to verify, we can positively conclude that we can calculate the HAL QCD potential from Lab frame NBS wave functions.

In conclusion, we confirm that potentials can be extracted at reasonable precision in the laboratory frame formalism of the HAL QCD method.
One lesson we learn is that we need more statistics than required in the conventional center-of-mass formalism.

\subsection{Scattering phase shifts} \label{subsect:numephase}
\begin{figure}[htbp]
  \begin{tabular}{cc}
    \begin{minipage}{0.5\hsize}
        \centering
        \includegraphics[width=\hsize,clip]{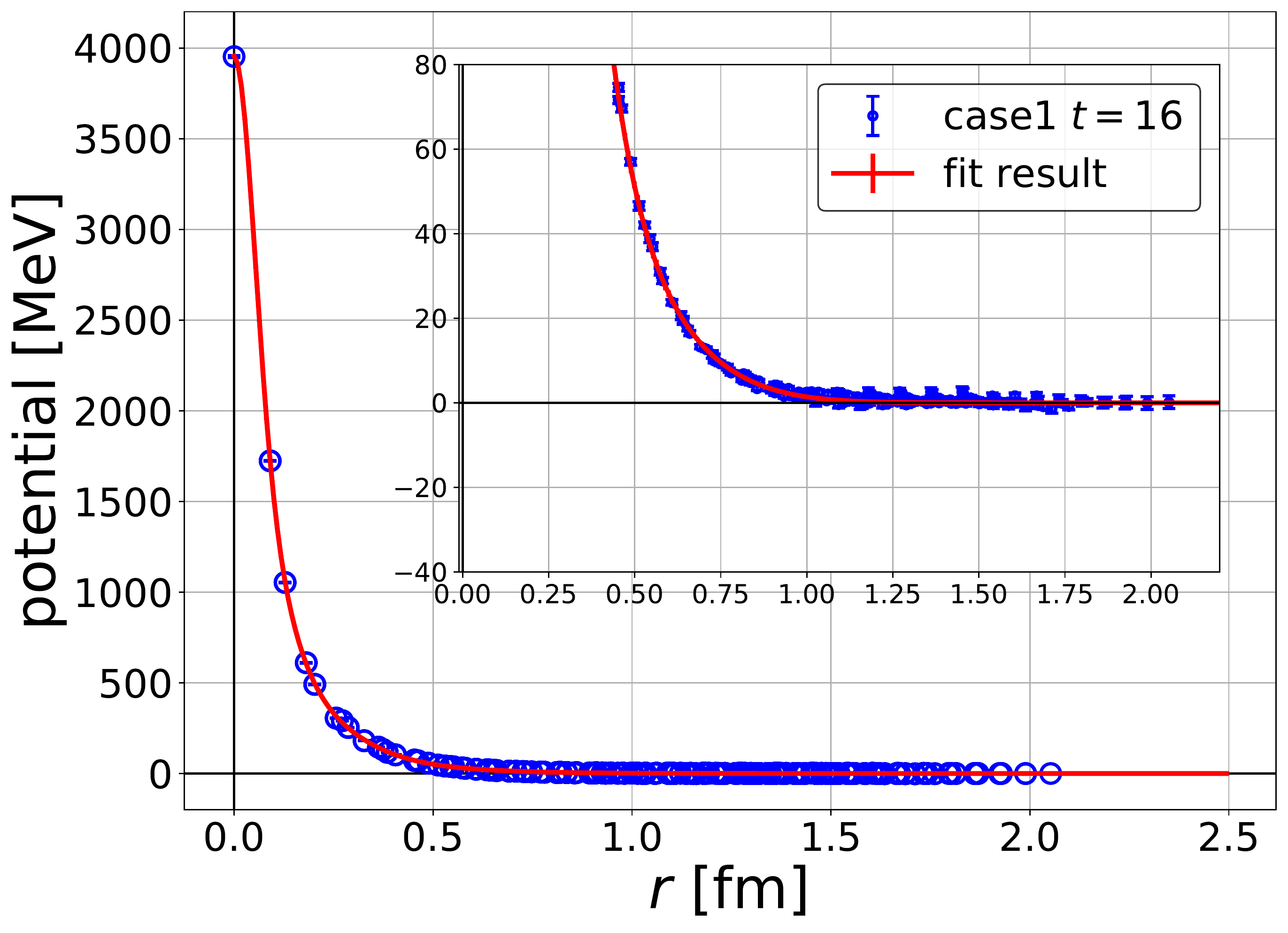}
    \end{minipage}
    &
    \begin{minipage}{0.5\hsize}
        \centering
        \includegraphics[width=\hsize,clip]{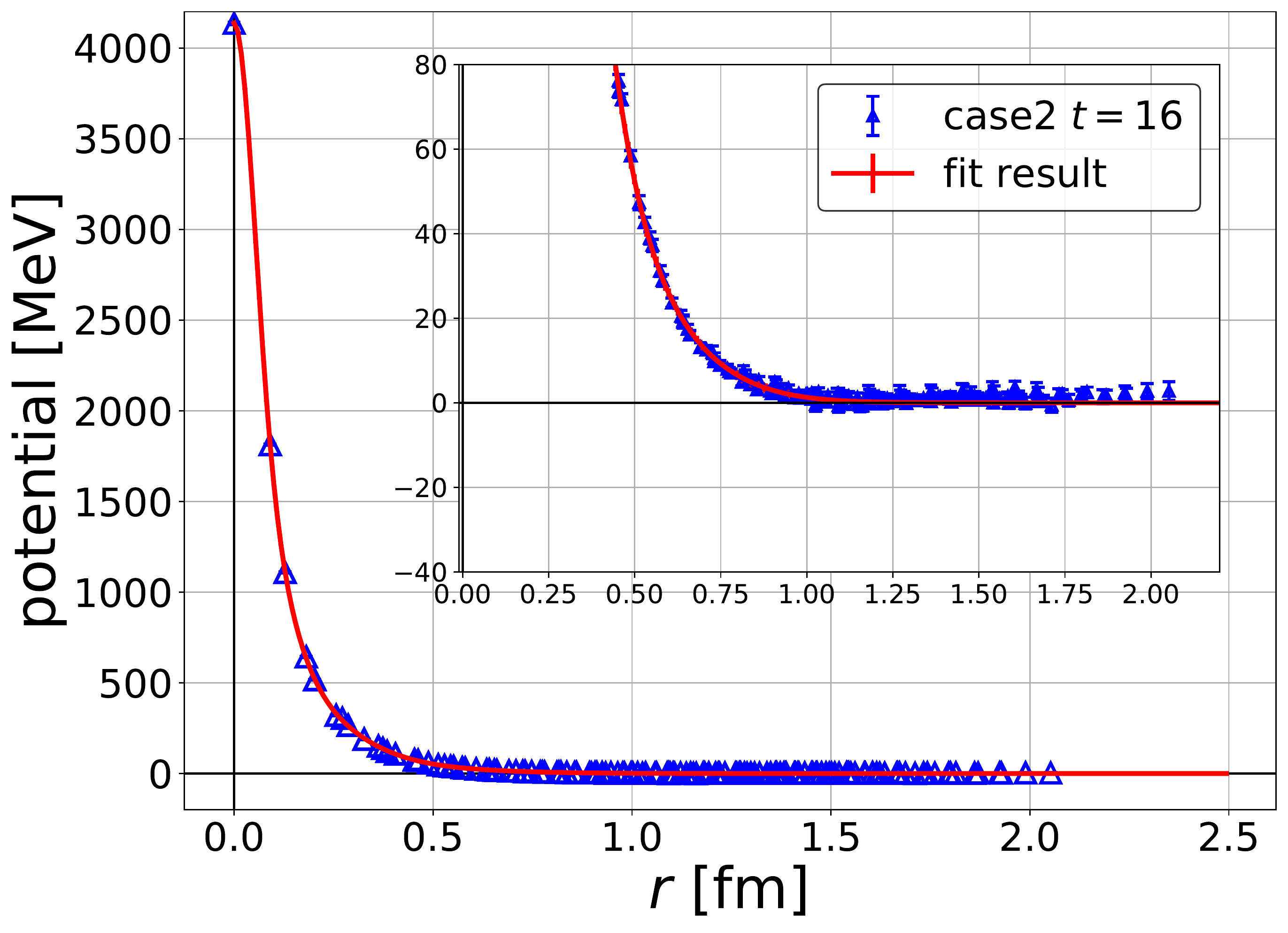}
    \end{minipage} \\
    \begin{minipage}{0.5\hsize}
        \centering
        \includegraphics[width=\hsize,clip]{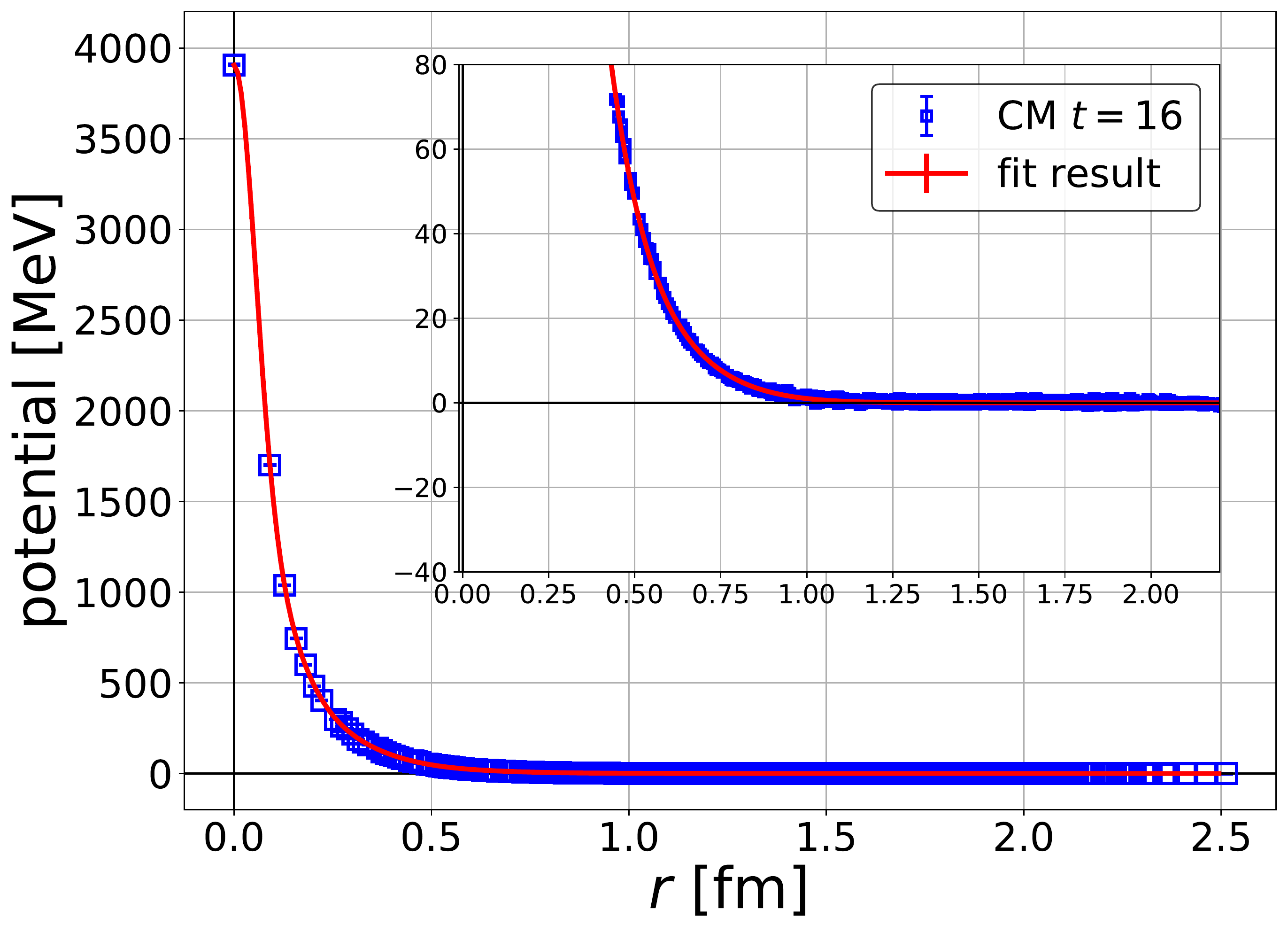}
    \end{minipage}
  \end{tabular}
  \caption{Fit results of effective LO potentials. Original data (blue points) and corresponding fit results (red lines) are shown.}
  \label{fig:i2pp_pot_fit}
\end{figure}

\begin{figure}[htbp]
  \centering
  \includegraphics[width=80mm,clip]{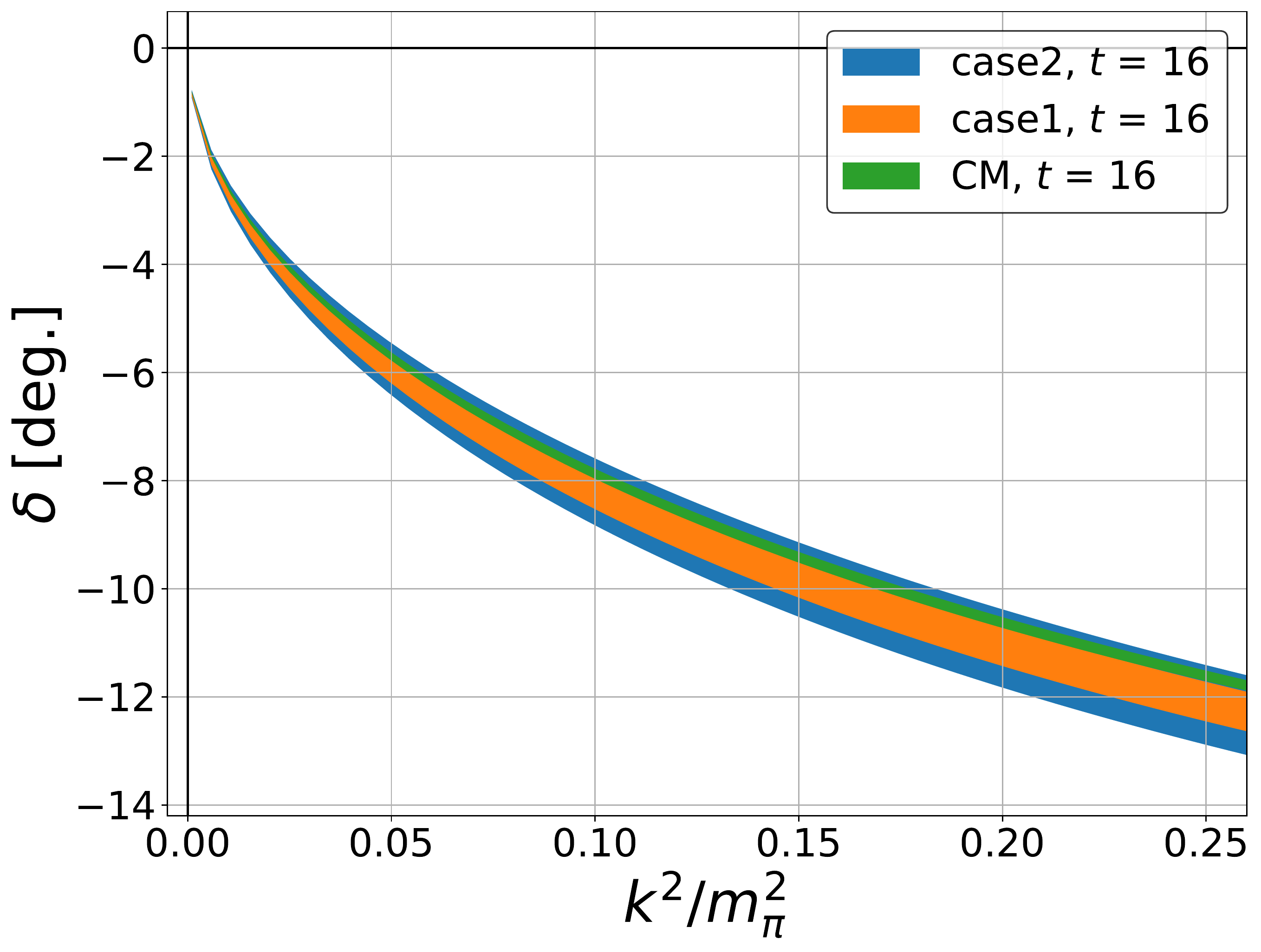}
  \caption{Scattering phase shifts obtained by effective LO potentials. Dark (light) color bands show statistical (systematical) errors.}
  \label{fig:phaseshift}
\end{figure}

{
To investigate a consistency between calculations in the Lab frame and the CM frame more precisely, let us compare behaviors of physical observables, such as scattering phase shifts $\delta_0(k)$ and $k \cot \delta_0(k)$.
%Then we calculate scattering phase shifts from the obtained potentials.
%We take the potentials at $X^4 = 15$ and systematic uncertainty coming from the time dependence is estimated by using data at $X^4 = 15 \pm 1$.
%{To calculate the phase shifts by solving the Schr\"odinger equation,}
We fit effective LO potentials with a sum of 4 Gaussians
\begin{equation}
  V(r) = a_0 e^{- (r/a_1)^2} + a_2 e^{- (r/a_3)^2} + a_4 e^{- (r/a_5)^2} + a_6 e^{- (r/a_7)^2},
  \label{eq:4Gauss}
\end{equation}
and solve the Schr\"odinger equation in infinte-volume.
%Systematic uncertainty coming from the time dependence is estimated by differences of data at $X^4 = 15 \pm 1$.
{Representative fit results and corresponding fit parameters are given in Fig.~\ref{fig:i2pp_pot_fit} and Tab.~\ref{tab:fit_params}, respectively.}
As seen from the table, values of $\chi^2/$d.o.f. are rather large, mainly due to data at short distances, which have smaller errors but
whose central values show scattered behaviors caused by higher partial waves.
Since fits in Fig.~\ref{fig:i2pp_pot_fit} looks reasonable in all cases,
we keep using the fitting function \eqref{eq:4Gauss}.
%{Phase shifts obtained from them are shown in Figure~\ref{fig:phaseshift}, respectively.}
%and resultant parameters are shown in Tab.~\ref{tab:fitparams}.
%Figure~\ref{} shows the resultant phase shifts.
We study systematic uncertainties of potential fits caused by discretization errors, whose details will be given in Appendix~\ref{appx:lab_hal_sys}.
Error bands in the following plots include both statistical and systematic errors.
As you can see from Figure~\ref{fig:phaseshift},
resultant scattering phase shifts obtained by the Lab frame calculations (blue and orange bands) are consistent with the conventional CM calculation (red band), as expected from the agreement of potentials.
%This observation indicates that the non-locality of the $I=2$ $\pi\pi$ potential in our setup is negligible, as mentioned in the previous study at heavier pion mass ($m_{\pi} \approx 870$ MeV)~\cite{Kawai:2017goq}.
%the potentials obtained by the Lab frame NBS wave functions (blue and orange bands) give consistent phase shifts with the conventional center-of-mass calculation (red band).
%This observation suggests that the non-locality of the $I=2$ $\pi\pi$ interaction does not affect  in our setup.
}

\begin{table}[htbp]
  \caption{Fit parameters of potentials at $t = X^4 = 16$.}
  \label{tab:fit_params}
  \begin{tabular}{c|cccccccc|c}
       & $a_0$ & $a_1$ & $a_2$ & $a_3$ & $a_4$ & $a_5$ & $a_6$ & $a_7$ & $\chi^2/$d.o.f. \\ \hline \hline
       %case1 & 0.543(14) & 1.306(60) & 0.275(18) & 2.48(20) & 0.100(26) & 4.47(31) & 0.898(39) & 0.7625(79) & 1.02 \\
       case1 & 0.5579(72) & 1.410(32) & 0.2551(58) & 2.86(10) & 0.052(11) & 5.24(24) & 0.953(22) & 0.7719(62) & 7.28 \\
       %case2 & 0.562(35) & 1.246(74) & 0.300(28) & 2.25(26) & 0.141(44) & 4.08(38) & 0.894(54) & 0.757(10) & 0.90 \\
       case2 & 0.5711(95) & 1.433(51) & 0.2618(75) & 2.88(16) & 0.055(19) & 5.15(44) & 1.014(34) & 0.7809(87) & 2.30 \\
       CM    & 0.474(10) & 1.589(75) & 0.206(12) & 3.04(19) & 0.045(15) & 5.16(34) & 1.070(36) & 0.813(11) & 2.57
  \end{tabular}
\end{table}

\begin{figure}[htbp]
  \begin{tabular}{cc}
    \begin{minipage}{0.5\hsize}
        \centering
        \includegraphics[width=0.85\hsize,clip]{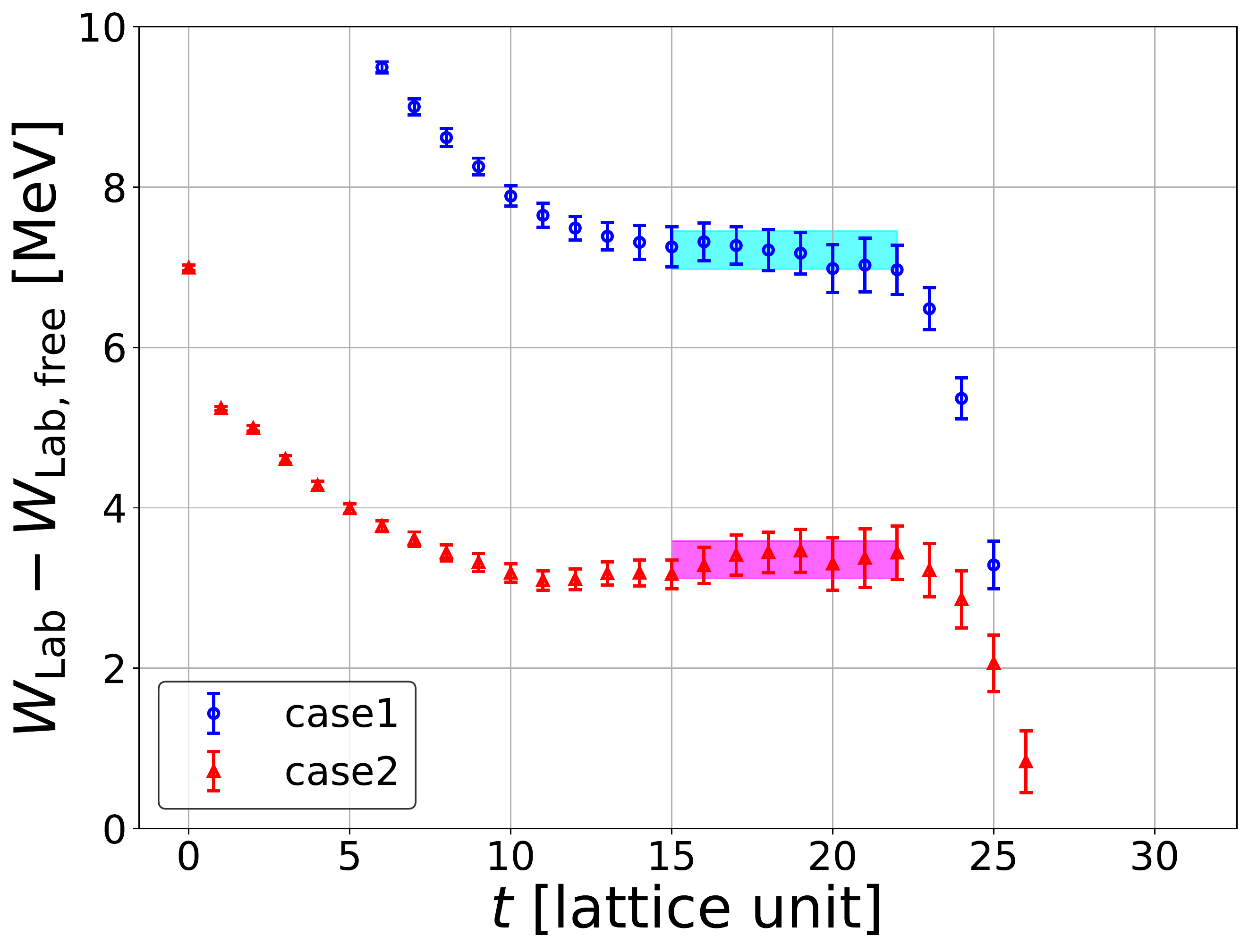}
    \end{minipage}
    &
    \begin{minipage}{0.5\hsize}
        \centering
        \includegraphics[width=0.85\hsize,clip]{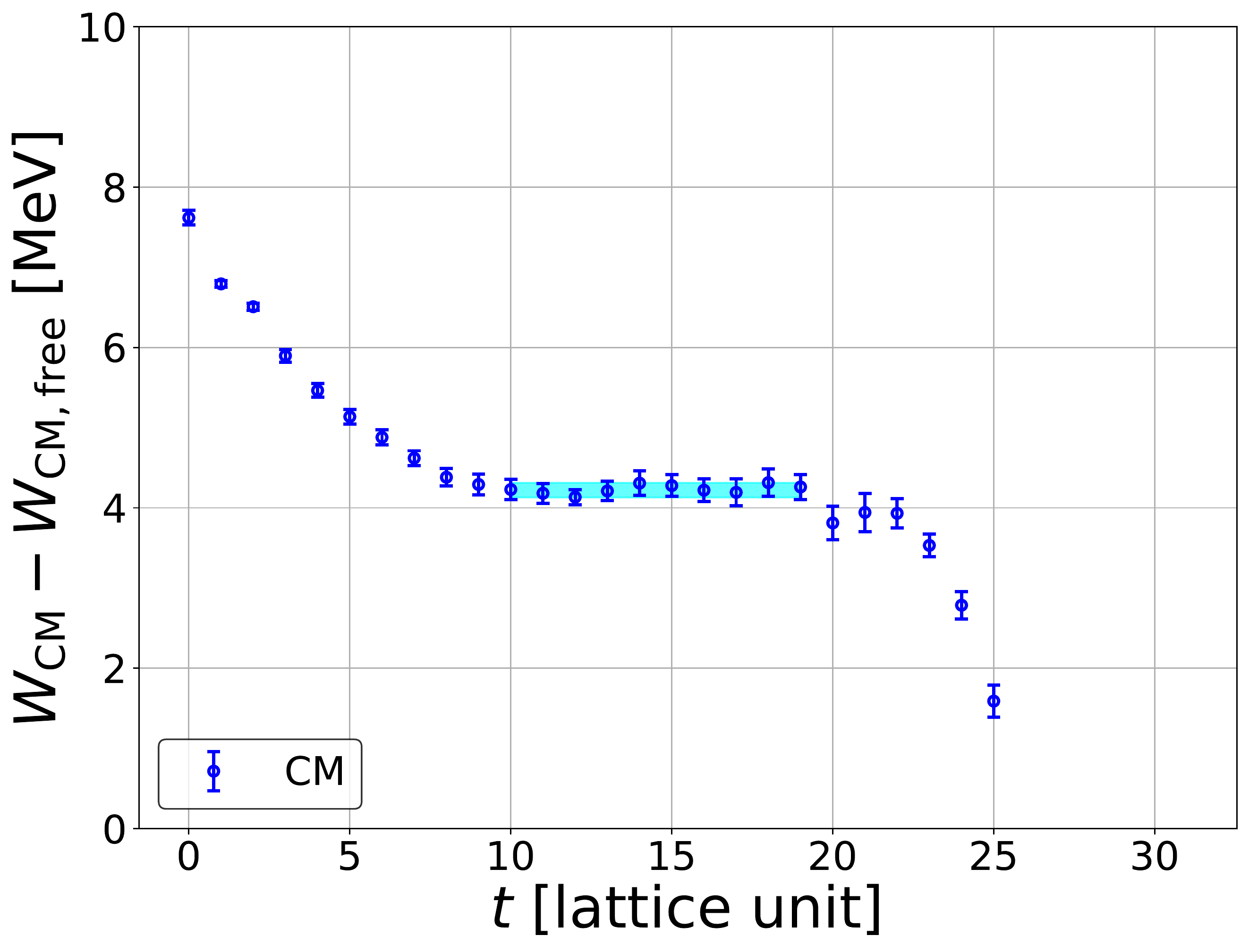}
    \end{minipage}
  \end{tabular}
  \caption{
	Extraction of energy shifts by single exponential fits in the laboratory frame (Left) and the center-of-mass frame (Right).
	Color bands show fit ranges and fit results with statistical errors.
	}
  \label{fig:fit_plateaus}
\end{figure}

\begin{figure}[htbp]
  \begin{tabular}{cc}
    \begin{minipage}{0.5\hsize}
        \centering
    \end{minipage}
    &
    \begin{minipage}{0.5\hsize}
        \centering
        \includegraphics[width=\hsize,clip]{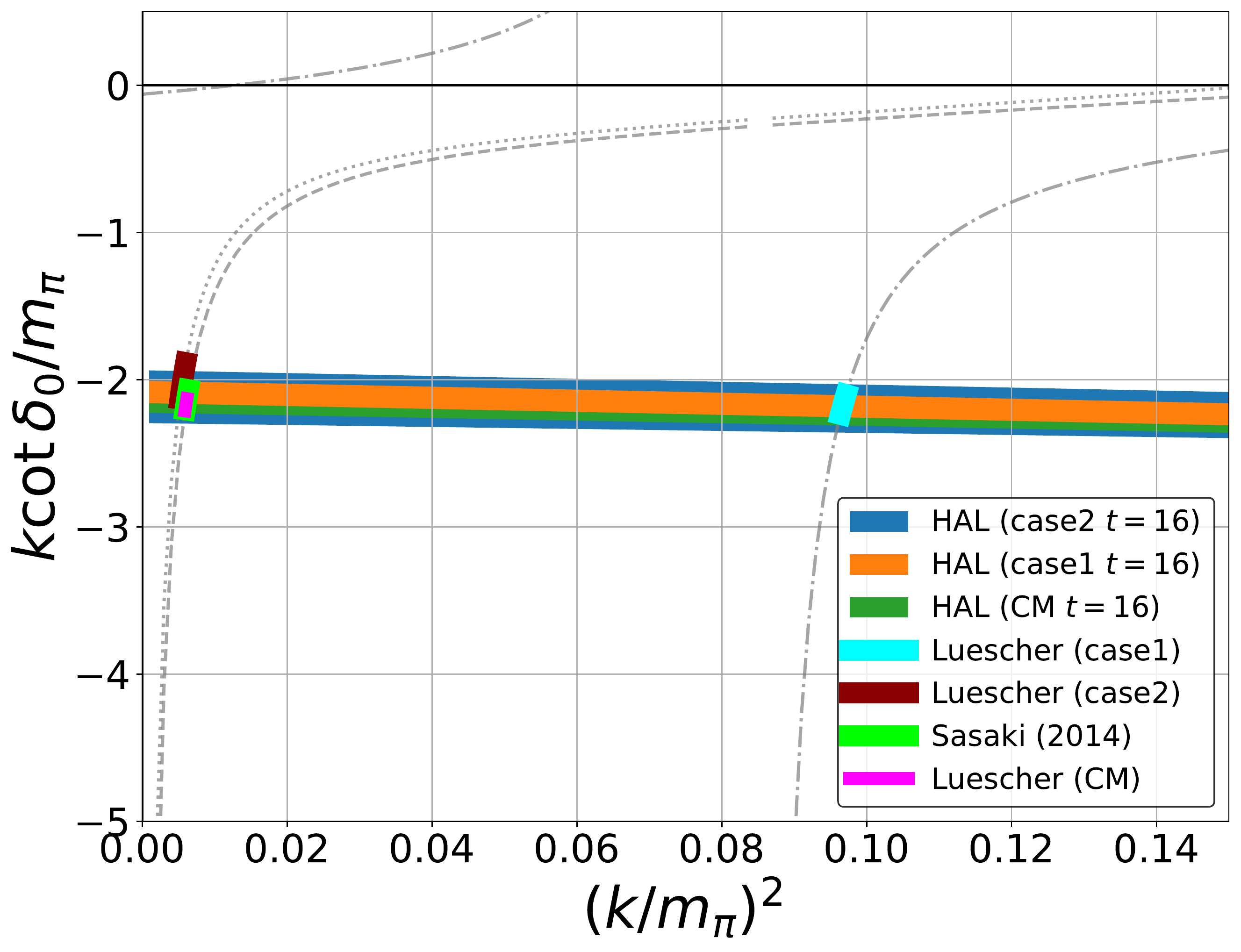}
    \end{minipage}
  \end{tabular}
  \caption{
  %(Left)
   A comparison of $k \cot \delta_0(k)$ between the HAL QCD method and the Luscher's method. Shown together are  lines of the Lushcer's formula.
  %(Right) Enlarged plot of the left figure.
  }
  \label{fig:kcotd}
\end{figure}

Finally, we compare our results with those obtained by the finite-volume method.
We extract ground state energies by a single exponential fit to the time dependence of the R-correlators, as shown in Fig.~\ref{fig:fit_plateaus}.
Energy levels are converted to the center-of-mass relative momentum $k_n$, to which the L\"uscher's formula is applied as
\begin{equation} \label{eq:lab_hal_luscherformula}
  k_n \cot \delta_0(k_n) = 4 \pi \frac{1}{\gamma_n L^3} \sum_{{\bf p} \in {\mathcal P}_{{\bf n}_{\rm total}}} \frac{1}{{\bf p}^2 - k_n^2},
\end{equation}
where
\begin{equation}
  {\mathcal P}_{{\bf n}_{\rm total}} = \{{\bf p} | {\bf p} = \frac{2 \pi}{L}\vec \gamma^{-1} ({\bf m}+ \frac{1}{2} {\bf n}_{\rm total}), \ {\bf m} \in {\bf Z}^3 \},
\end{equation}
with a short-hand notation $\vec \gamma^{-1}{\bf n} = \gamma^{-1} {\bf n}_{\parallel} + {\bf n}_{\perp}$.
Extracted values of $k\cot \delta_0(k)$ are plotted in Fig.~\ref{fig:kcotd} (Right), together with a result in the literature~\cite{Sasaki:2013vxa}.
As seen in the figure, we confirm that phase shifts obtained by the HAL QCD method are consistent with those by the finite-volume method.
It also implies that the LO approximation is valid in the energy region we consider here, since the finite-volume method is free from systematics associated with the derivative expansion.
%Note that the energy levels extracted here may suffer from "fake plateau" since we just employ naive extraction, the exponential fit.

%%%%% result end %%%%%

%%%%% conclusion %%%%%
\section{Summary} \label{sect:summary}
In this paper, we propose a theoretical framework to calculate HAL QCD potentials from NBS wave functions in laboratory frames,
and perform a first numerical calculation in the $I=2$ $\pi \pi$ system at $m_{\pi} \approx 700$ MeV.
We calculate effective LO potentials from NBS wave functions with total momenta $P = (0,0,n) \quad (n = 0,1,2)$.
While larger statistical fluctuations and non-smooth behaviors, which are probably originated  from higher order numerical derivatives and the reduced rotational symmetry, have been observed in laboratory frames ($n=1,2$),
potentials in all cases ($n=0,1,2$) are repulsive and agree with each other except small deviations at short distances.
Resultant phase shifts $\delta_0(k)$ and $k \cot \delta_0(k)$ with $n=1,2$ are consistent with those obtained not only by the conventional center-of-mass calculation ($n=0$) but also by the finite-volume method.
In conclusion, we confirm that the laboratory frame formalism works in practice to extract scattering phase shifts in lattice QCD.
As already mentioned in Sect.~\ref{sect:intro},  it enlarges applicabilities and opens new possibilities for the HAL QCD method,
such as determinations of higher-order terms in the derivative expansion of non-local potentials and
extractions of potentials for systems having same quantum numbers with a vacuum state,

We finally discuss some issues for a use of laboratory frames in the HAL QCD method.
As already mentioned,
statistical fluctuations are larger in laboratory frames, probably due to larger energy of states with non-zero momenta and
higher order time derivatives necessary for the time dependent method.
While a number of statistical sampling required for meaningful results is manageably small for the $I=2$  $\pi\pi$ system,
it may drastically increase for more complicated systems including quark-antiquark pair creations and annihilations.
Non-smooth behaviors of potentials, caused by  reduced symmetries in laboratory frames,
may be cured by the partial wave decomposition technique~\cite{Miyamoto:2019jjc}, though the technique is restricted to the center-of-mass system at this moment.
Since larger non-zero momenta may cause lager discretization errors through violations of continuum dispersion relations,
we should always check a validity of the continuum dispersion relation. In addition, it is better to
compare different normalizations of R-correlators for potentials,
as discussed in Appendix~\ref{appx:lab_hal_sys}.

%%%%% conclusion end %%%%%

%%%%% acknowledgement %%%%%
\begin{acknowledgements}
The authors thank members of the HAL QCD Collaboration for fruitful discussions.
We thank the PACS-CS Collaboration [38] and ILDG/ JLDG [46] for providing their configurations.
The numerical simulation in this study is performed on the Oakforest-PACS in Joint Center for Advanced High Performance Computing (JCAHPC).
The framework of our numerical code is based on Bridge++ code~\cite{Ueda:2014rya} and its optimized version for the Oakforest-PACS by Dr. I. Kanamori~\cite{Kanamori:2018hwh}.
Y.~A. is supported in part by the Japan Society for the Promotion of Science (JSPS).
S.~A. is supported in part by he Grant-in-Aid of the MEXT for Scientific Research (Nos. JP16H03978, JP18H05236).

\end{acknowledgements}
%%%%% acknowledgement end %%%%%

%%%%% Appendix %%%%%
\appendix
\section{Systematic uncertainties} \label{appx:lab_hal_sys}

In this appendix, we estimate systematic uncertainties on extractions of $I=2$  S-wave $\pi\pi$ scattering phase shifts
in laboratory frames.
Concretely, we investigate a normalization dependence and an $X^4$ dependence of the potential extraction with non-zero total momenta.

For the former investigation, we consider an alternative time-dependent method using R-correlators normalized differently as
\begin{eqnarray}
	R_{\bf P=e_z} ({\bf x},x^4,X^4) &=& \frac{F_{\pi^+ \pi^+, {\bf e_z}}({\bf x},x^4,X^4)}{F_{\pi^+,0}(X^4)^2} \\
	R_{\bf P=2e_z} ({\bf x},x^4,X^4) &=& \frac{F_{\pi^+ \pi^+, {\bf 2e_z}}({\bf x},x^4,X^4)}{F_{\pi^+,0}(X^4)^2},
\end{eqnarray}
where $F_{\pi^+,0}(X^4) = \sum_{\bf x,y} \langle \pi^+({\bf x},X_4) \pi^-({\bf y},0) \rangle$ is a pion 2-pt function with {\it zero} momentum.
Building blocks of potentials are thus modified to
\begin{eqnarray}
  G({\bf x},x^4,X^4) &=& \left( (\partial_{X^4} - 2m)^2 - {\bf P}^2 \right) R_{\bf P}({\bf x},x^4,X^4), \\
  E({\bf x},x^4,X^4) &=& \frac{1}{4m} \left[ \partial_{X^4}^2 - 4m \partial_{X^4} - {\bf P}^2 \right] G({\bf x},x^4,X^4),
  \label{eq:def_Em}\\
  L_{\perp}({\bf x},x^4,X^4) &=& \nabla_{\perp}^2 G({\bf x},x^4,X^4),\\
  L_{\parallel}({\bf x},x^4,X^4) &=& \left( - (\partial_{X^4} - 2m) \nabla_{\parallel} + i {\bf P} \partial_{x^4} \right)^2 R_{\bf P}({\bf x},x^4,X^4),
\end{eqnarray}
which should be compared with eqs.~\eqref{eq:def_G} --- \eqref{eq:lapC}.
Using these, the LO potential is constructed as
\begin{equation} \label{eq:effLOpotM}
  V^{\rm LO}_{x^{*4}=0}({\bf x}_{\perp}) = \left. \frac{\left(L_{\perp} + L_{\parallel} + mE \right)({\bf x},x^4,X^4)}{mG({\bf x},x^4,X^4)} \right|_{x^4=0, {\bf x}_{\parallel}=0},
\end{equation}
which is expected to agree with the one in \eqref{eq:effLOpot} within systematics errors.
Therefore
we can estimate the systematics from a difference between two potentials with different normalizations.

For the latter, we simply compare potentials at $X^4 = 16 \pm 1$.

%By comparing all results, we estimate final systematic uncertainty. % by the maximum and minimum values of them.
In the following, we show both dependences,  and present an estimation of uncertainties on scattering phase shifts.

\subsection{Normalization dependence}
\begin{figure}[htbp]
  \begin{tabular}{cc}
    \begin{minipage}{0.5\hsize}
        \centering
        \includegraphics[width=0.95\hsize,clip]{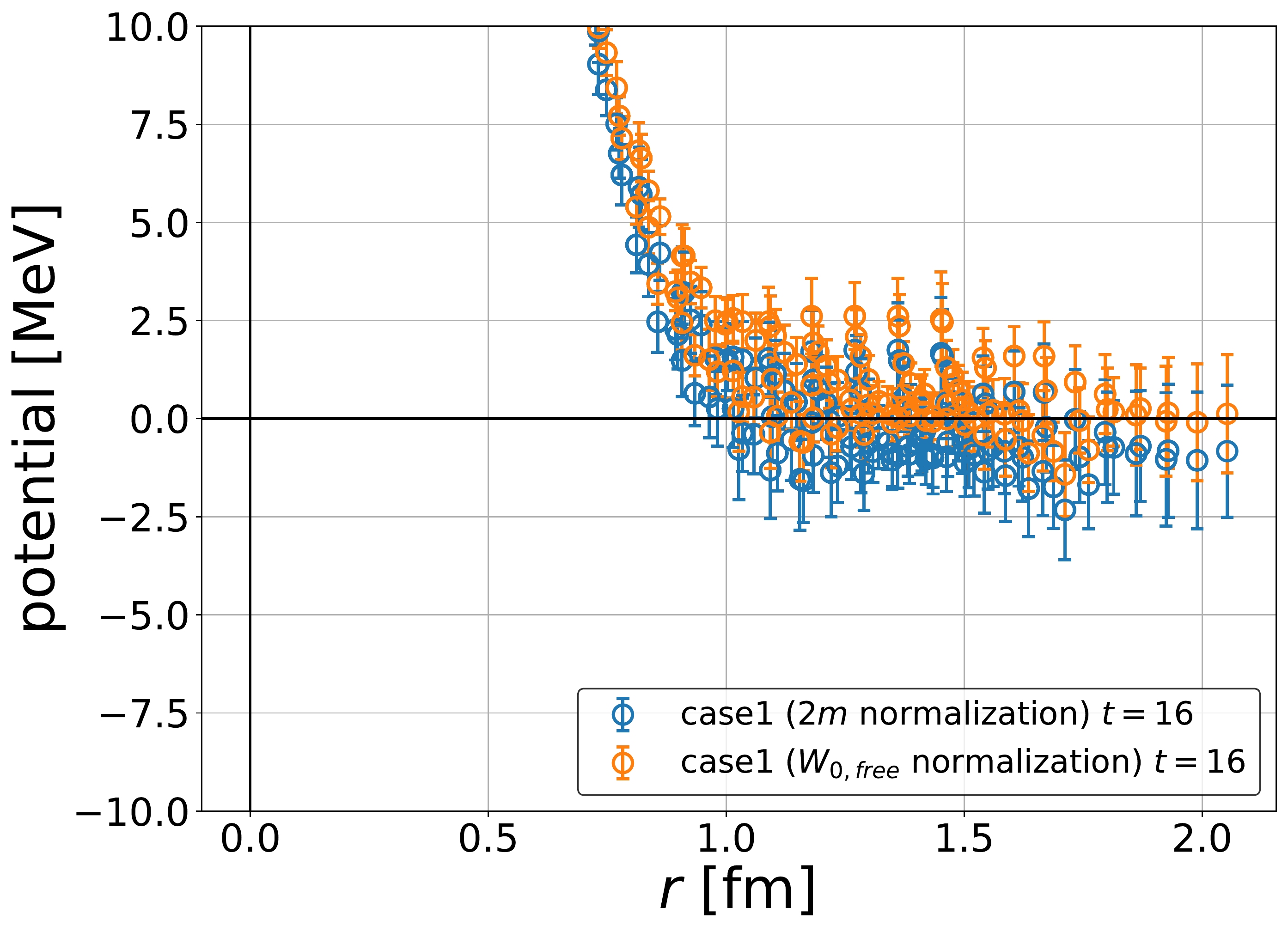}
    \end{minipage}
    &
    \begin{minipage}{0.5\hsize}
        \centering
        \includegraphics[width=0.95\hsize,clip]{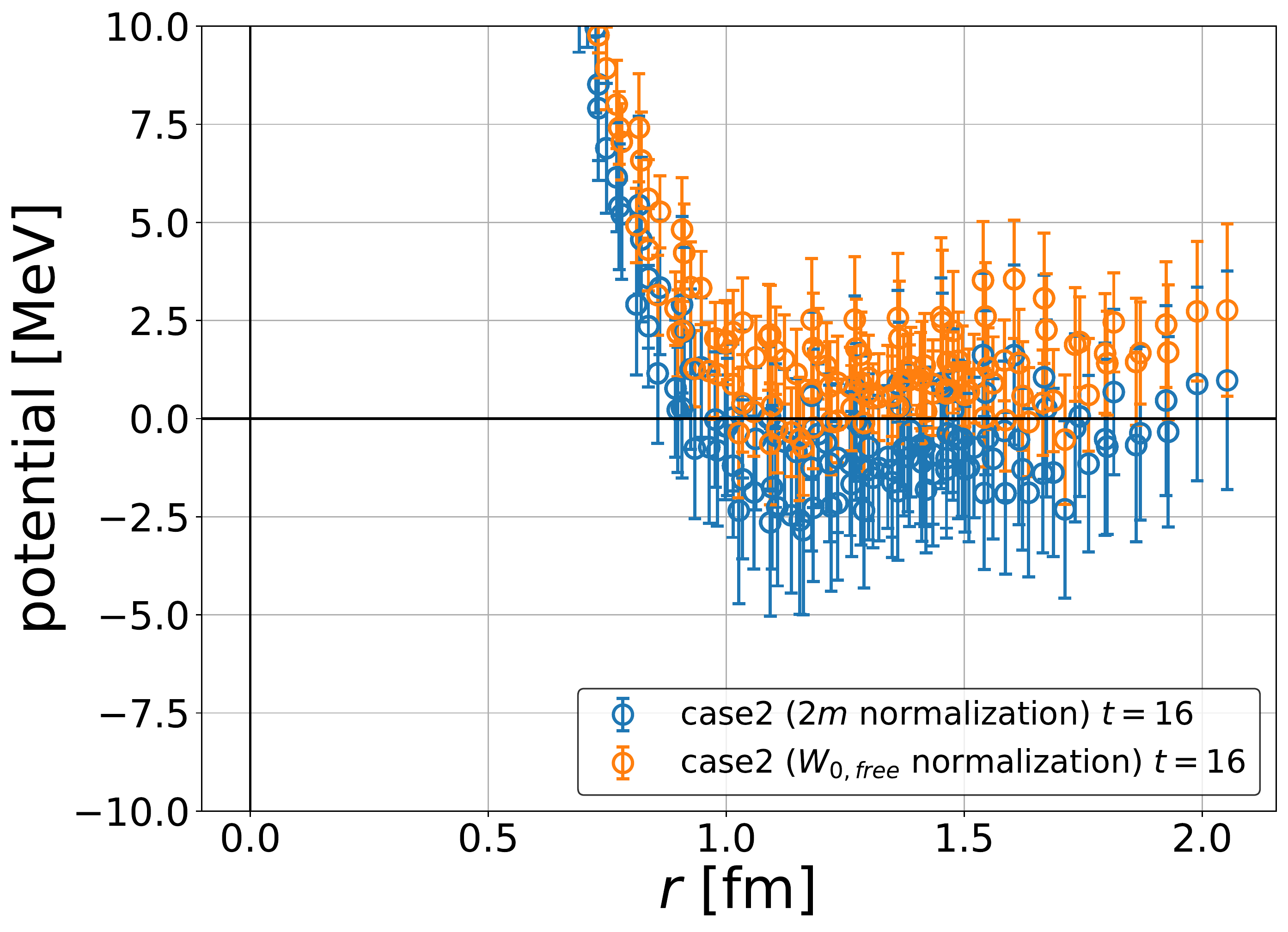}
    \end{minipage}
  \end{tabular}
  \caption{
	A normalization dependence of the potential in the case 1 (Left) and the case 2 (Right).
  %We also show  ?? the center-of-mass result ?? (black diamons) for  comparison.
  }
  \label{fig:appx:pot_normdep}
\end{figure}

\begin{figure}[htbp]
  \begin{tabular}{cc}
    \begin{minipage}{0.5\hsize}
        \centering
        \includegraphics[width=0.95\hsize,clip]{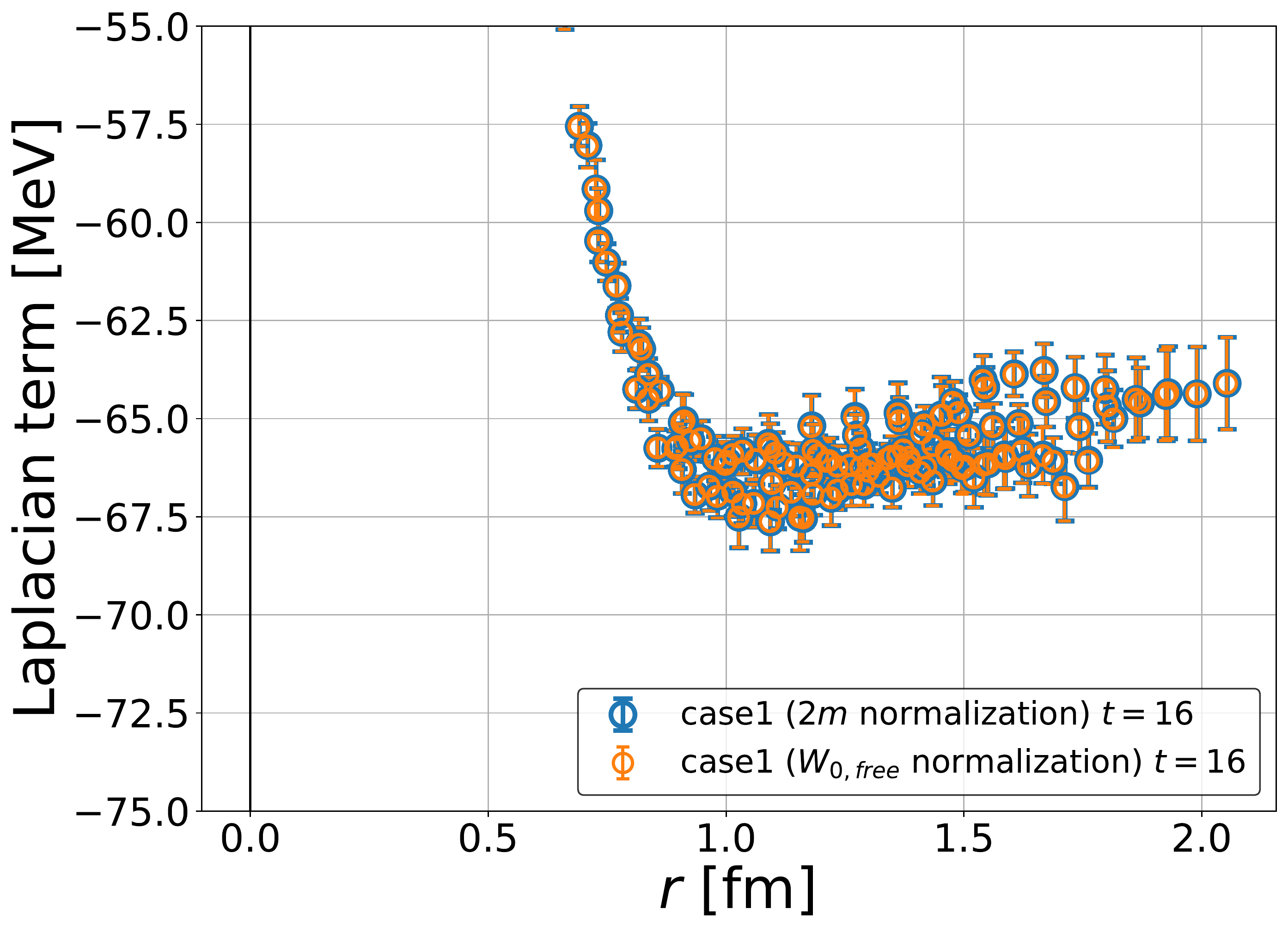}
    \end{minipage}
    &
    \begin{minipage}{0.5\hsize}
        \centering
        \includegraphics[width=0.95\hsize,clip]{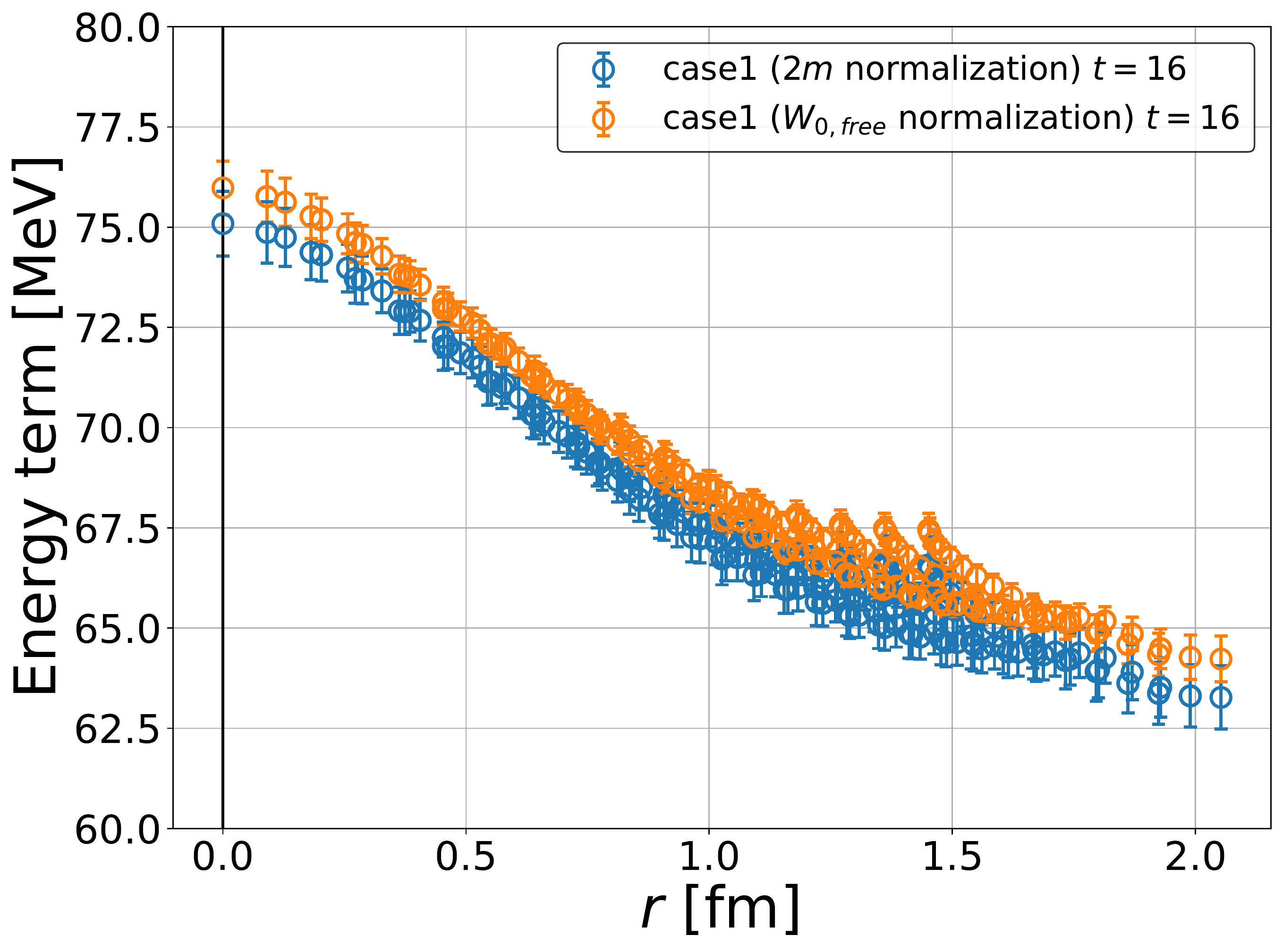}
    \end{minipage}
  \end{tabular}
  \caption{
	A normalization dependence of the Laplacian  term (Left) and the energy term (Right) in the case 1.
  }
  \label{fig:appx:potdecomp_normdep_case1}
\end{figure}

\begin{figure}[htbp]
  \begin{tabular}{cc}
    \begin{minipage}{0.5\hsize}
        \centering
        \includegraphics[width=0.95\hsize,clip]{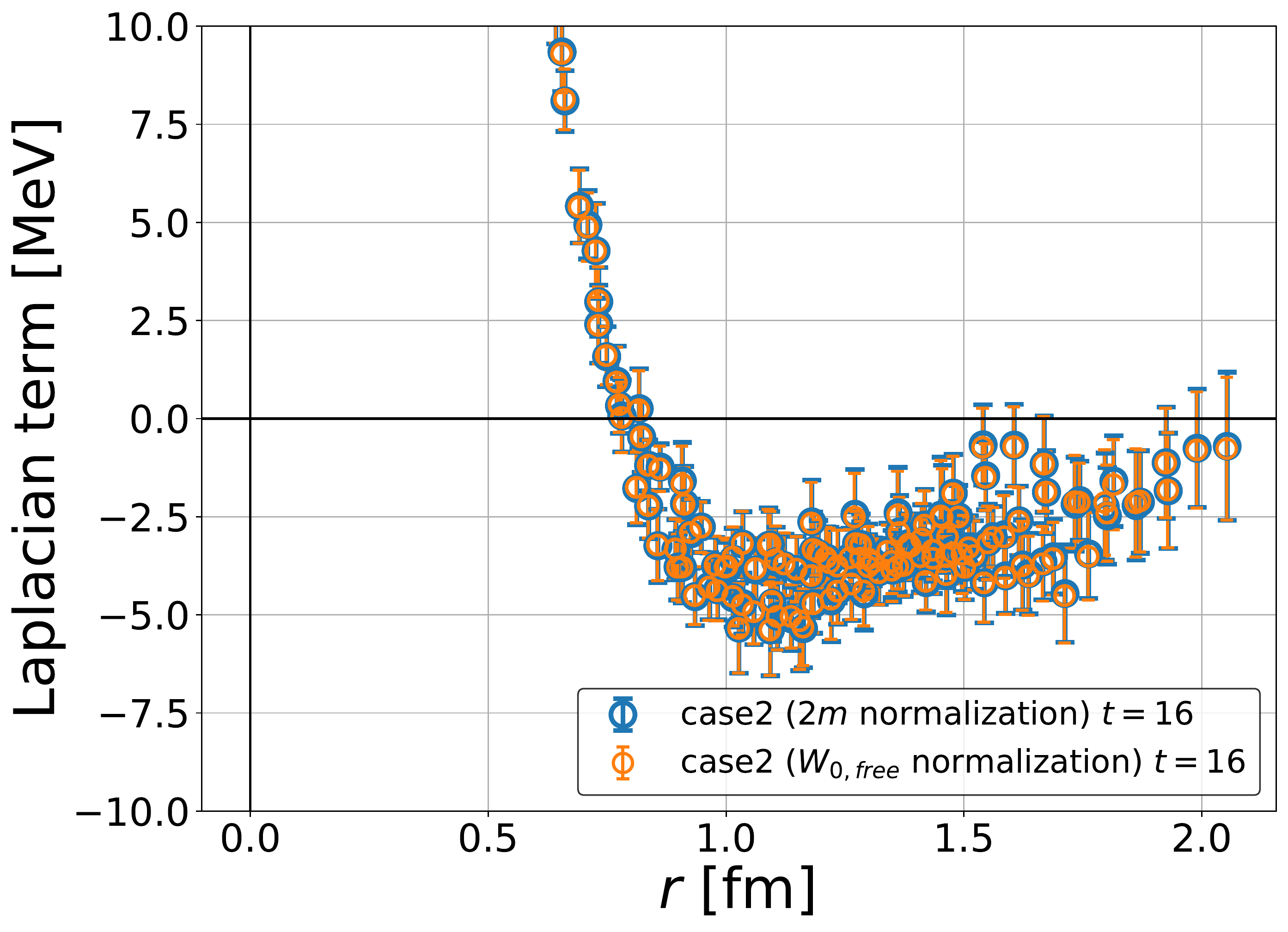}
    \end{minipage}
    &
    \begin{minipage}{0.5\hsize}
        \centering
        \includegraphics[width=0.95\hsize,clip]{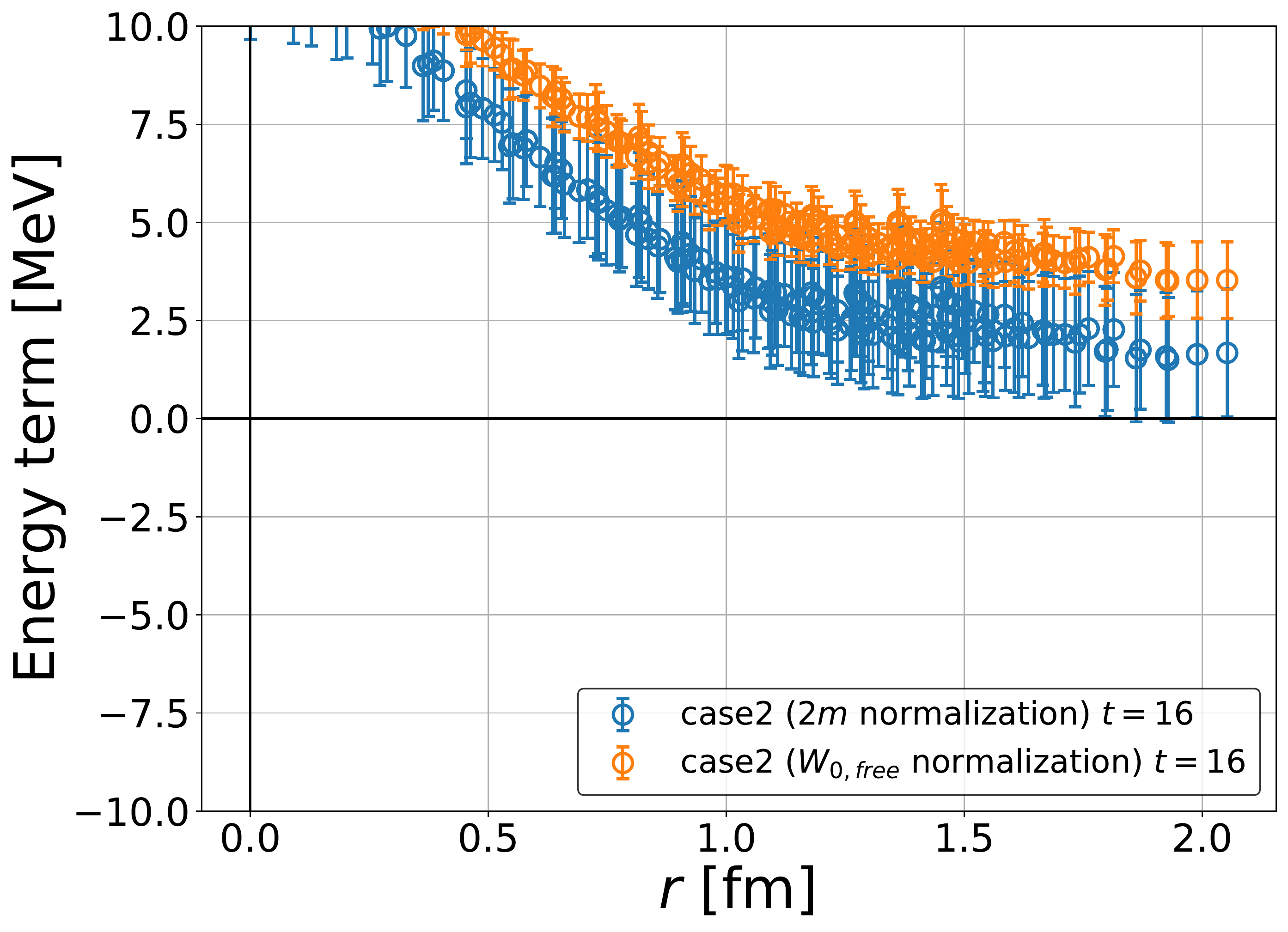}
    \end{minipage}
  \end{tabular}
  \caption{
	A normalization dependence of the Laplacian term (Left) and the energy term (Right) in the case 2.
  }
  \label{fig:appx:potdecomp_normdep_case2}
\end{figure}

Figure~\ref{fig:appx:pot_normdep} shows the normalization dependence of effective LO potentials,
the $W_{0,{\rm free}}$ normalization (orange) defined in eq.~\eqref{eq:effLOpot}  and the $2m$ normalization (blue) in eq.~\eqref{eq:effLOpotM},
with non-zero total momenta, ${\bf P = e_z}$ (Left) and  ${\bf P = 2 e_z}$ (Right).
We observe a slight systematic downward shift of central values for the LO potential with the $2m$ normalization,
though differences are comparable with sizes of statistical errors,
and these shifts mainly come from the energy term, as seen in Figure~\ref{fig:appx:potdecomp_normdep_case1} and \ref{fig:appx:potdecomp_normdep_case2}.
Since an implicit estimation of $k_m^2$ in the energy term, \eqref{eq:def_E} or \eqref{eq:def_Em}, relies on the continuum dispersion relation,
and involves a discretized 2nd time derivative $\partial_{X_4}^2$,
we suspect that these shifts are caused by finite lattice spacing effects for a larger energy of moving particles,
and thus are regarded as discretization errors associated with laboratory frame calculations.

Since potentials should become zero at long distances and indeed are consistent with zero within statistical errors,
we exclude data at such longer distances and
employ data at  $r < 13$(1.17[fm]) for the fit of potentials,  \eqref{eq:4Gauss}, in order to reduce
statistical and systematic fluctuations of potentials at longer distances as much as possible.

\begin{figure}[htbp]
  \begin{tabular}{cc}
    \begin{minipage}{0.5\hsize}
        \centering
        \includegraphics[width=0.95\hsize,clip]{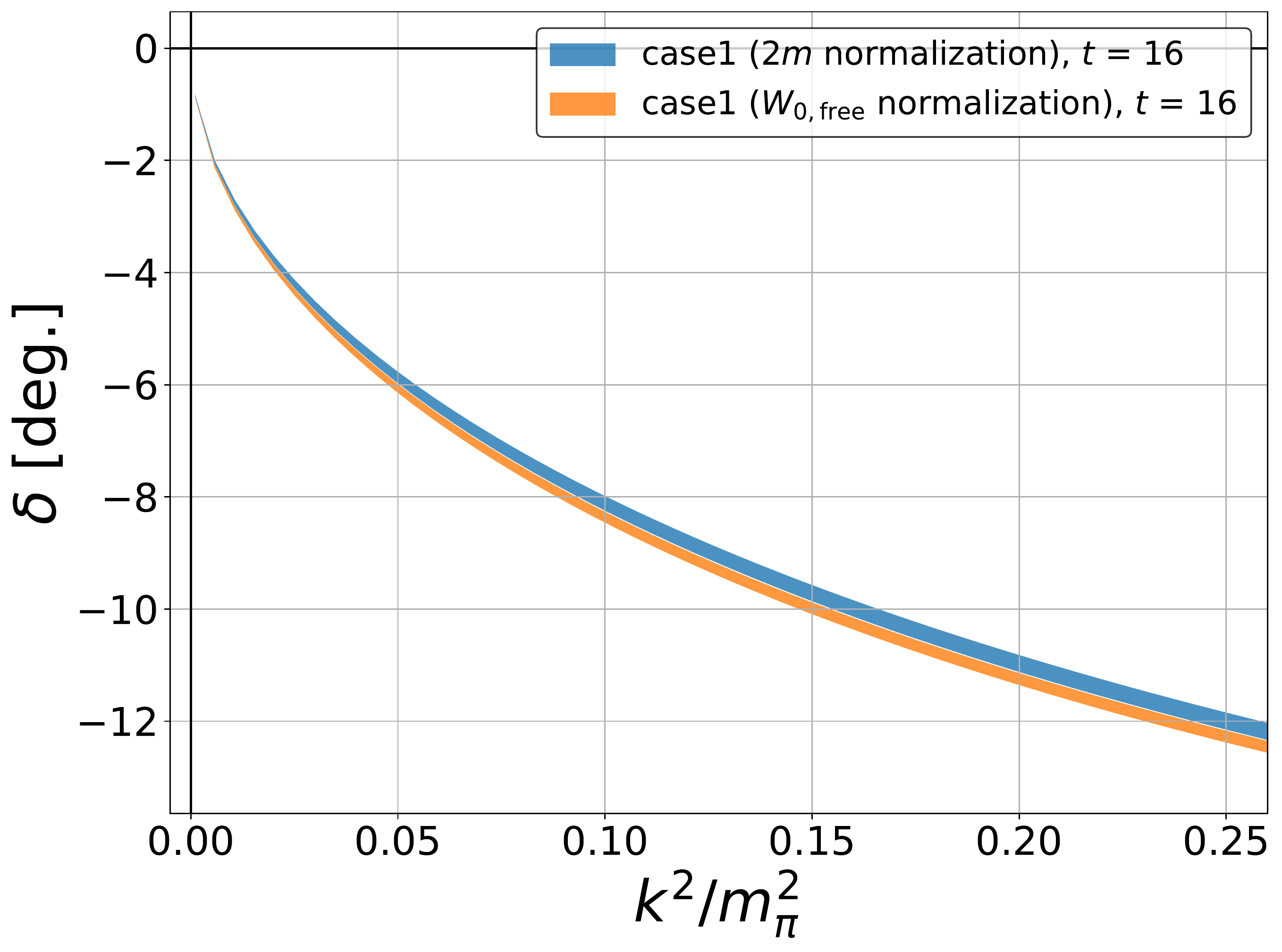}
    \end{minipage}
    &
    \begin{minipage}{0.5\hsize}
        \centering
        \includegraphics[width=0.95\hsize,clip]{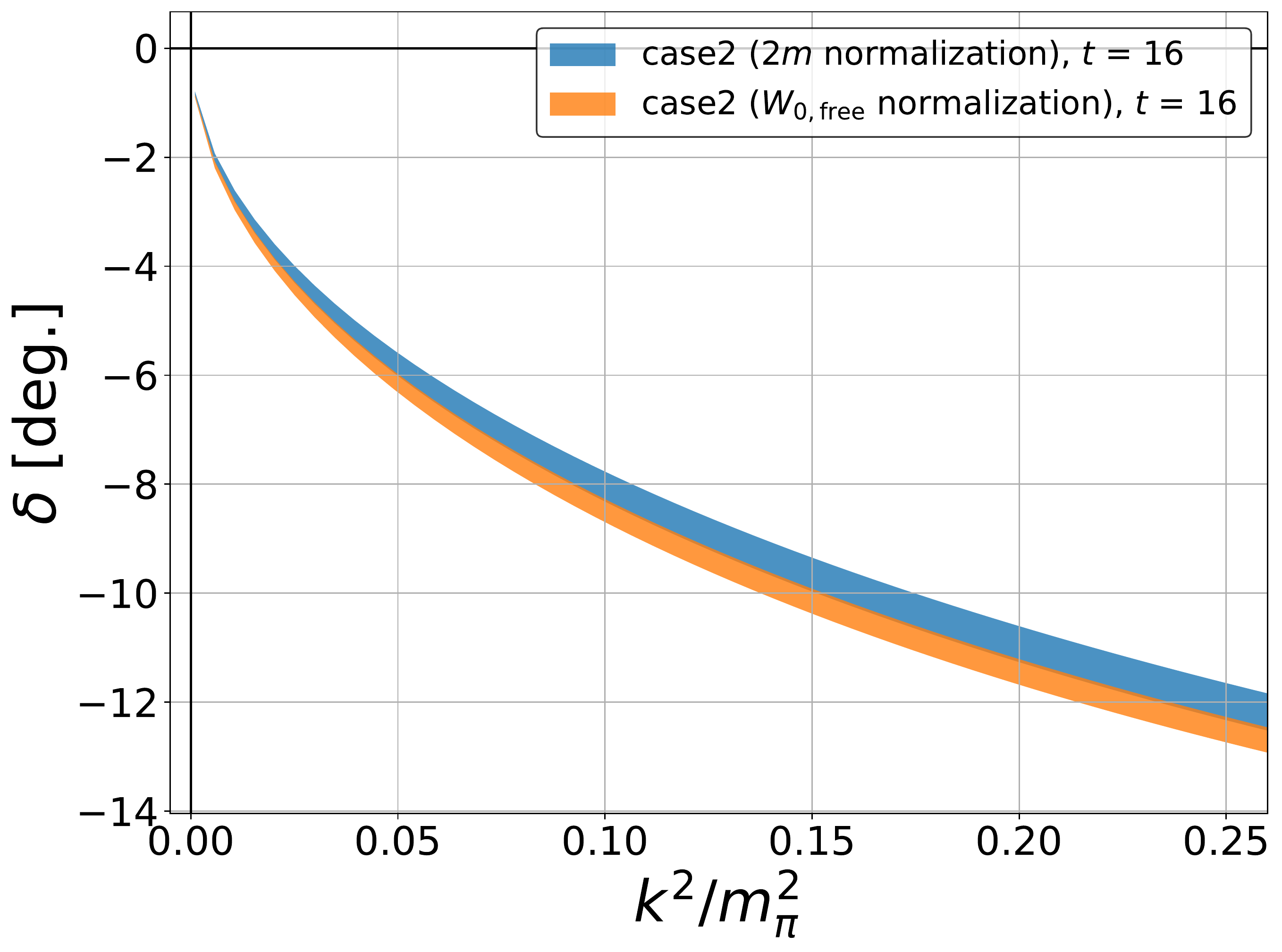}
    \end{minipage}
  \end{tabular}
  \caption{
	A normalization dependence of scattering phase shifts at $X^4 = 16$ in the case 1 (Left) and the case 2 (Right).
  }
  \label{fig:appx:phaseshift_normdep}
\end{figure}

Fig.~\ref{fig:appx:phaseshift_normdep} shows scattering phase shifts as a function of $k^2/m_\pi^2$
in two normalizations.
We observe a slight difference between the two, which is therefore taken into account for
our estimation of systematic uncertainties.

\subsection{Time dependence}
%\section{$X^4$ dependence}

\begin{figure}[htbp]
  \begin{tabular}{cc}
    \begin{minipage}{0.5\hsize}
        \centering
        \includegraphics[width=0.95\hsize,clip]{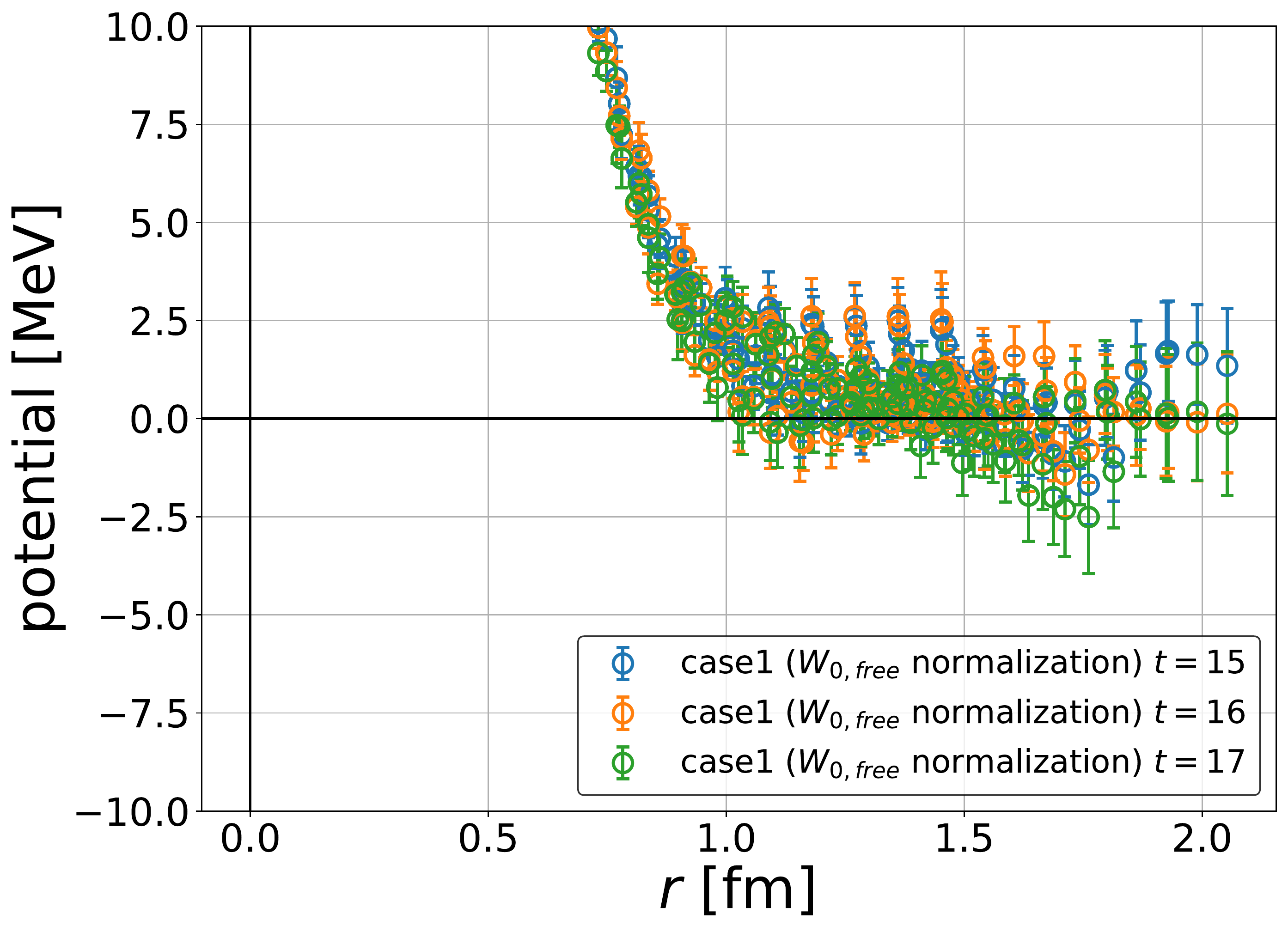}
    \end{minipage}
    &
    \begin{minipage}{0.5\hsize}
        \centering
        \includegraphics[width=0.95\hsize,clip]{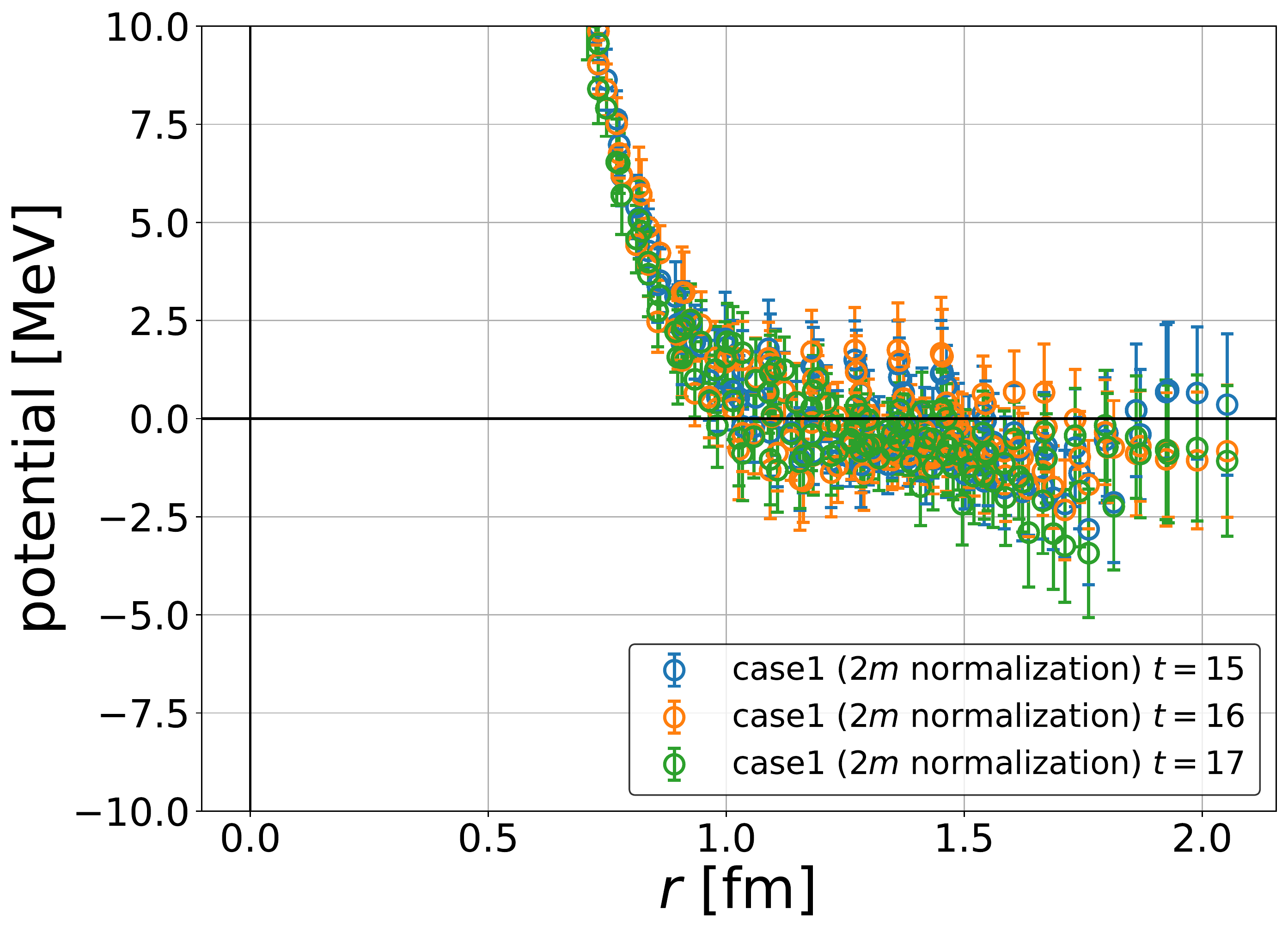}
    \end{minipage}
  \end{tabular}
  \caption{
	The $X^4$ dependence of potentials using $W_{0,{\rm free}}$ normalization (Left) and $2m$ normalization (Right) in the case 1.
	%The center-of-mass result at $X^4 = 16$ is shown as {\bf black diamonds} for a comparison.
  }
  \label{fig:appx:pot_tdep_case1}
\end{figure}

\begin{figure}[htbp]
  \begin{tabular}{cc}
    \begin{minipage}{0.5\hsize}
        \centering
        \includegraphics[width=0.95\hsize,clip]{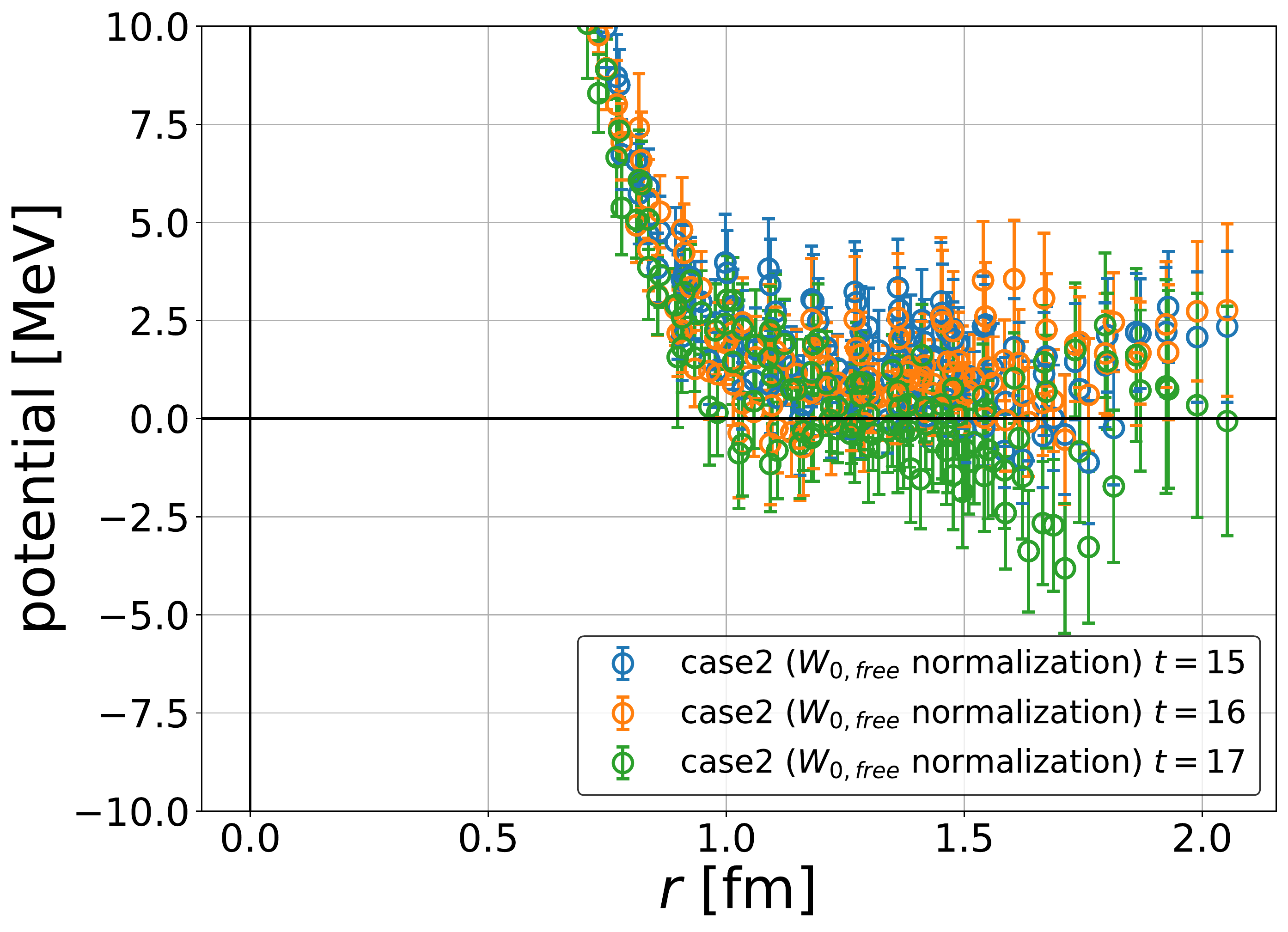}
    \end{minipage}
    &
    \begin{minipage}{0.5\hsize}
        \centering
        \includegraphics[width=0.95\hsize,clip]{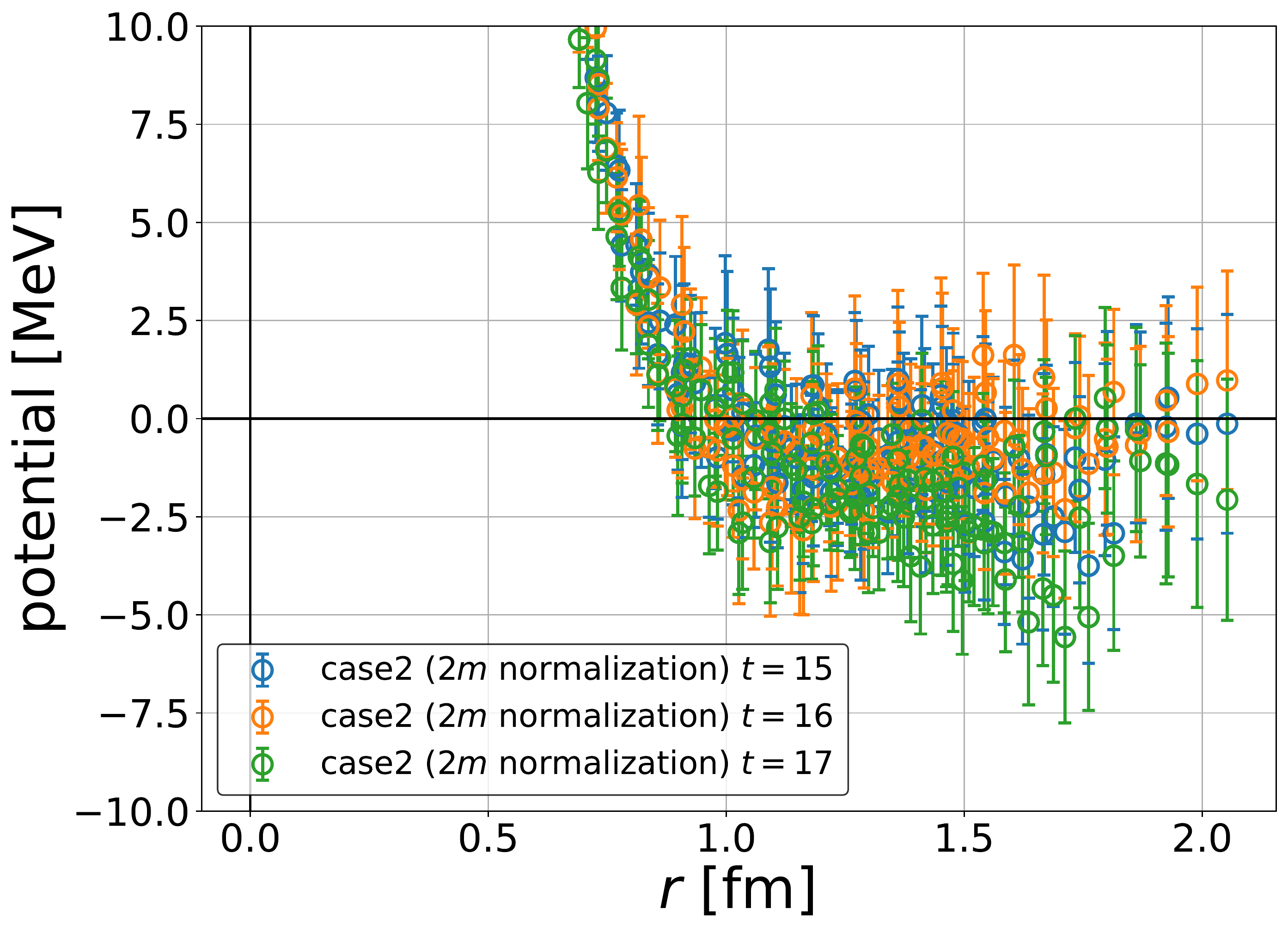}
    \end{minipage}
  \end{tabular}
  \caption{
	Same as Fig.~\ref{fig:appx:pot_tdep_case1} in the case 2.
	}
  \label{fig:appx:pot_tdep_case2}
\end{figure}

\begin{figure}[htbp]
  \begin{tabular}{cc}
    \begin{minipage}{0.5\hsize}
        \centering
        \includegraphics[width=0.95\hsize,clip]{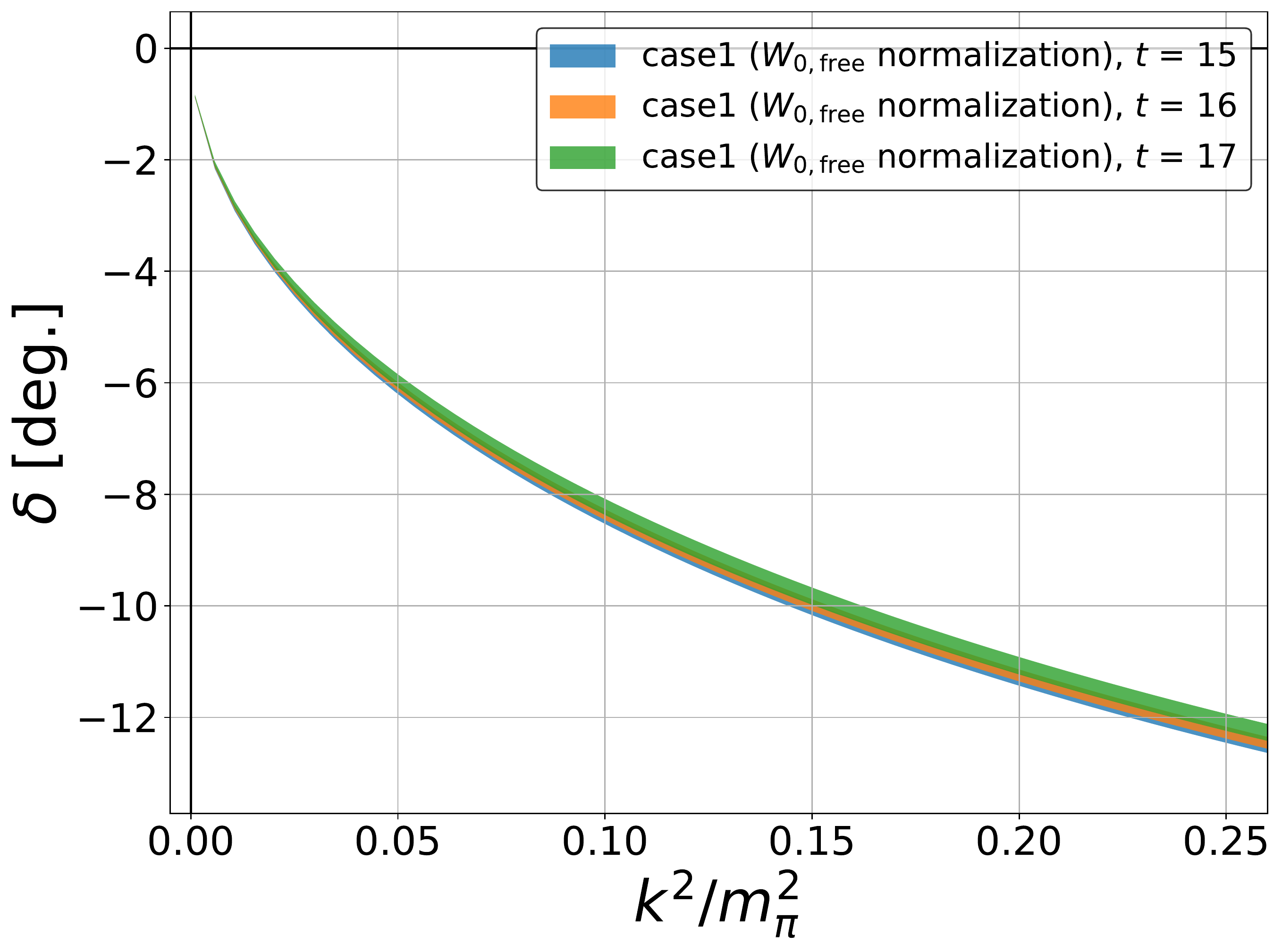}
    \end{minipage}
    &
    \begin{minipage}{0.5\hsize}
        \centering
        \includegraphics[width=0.95\hsize,clip]{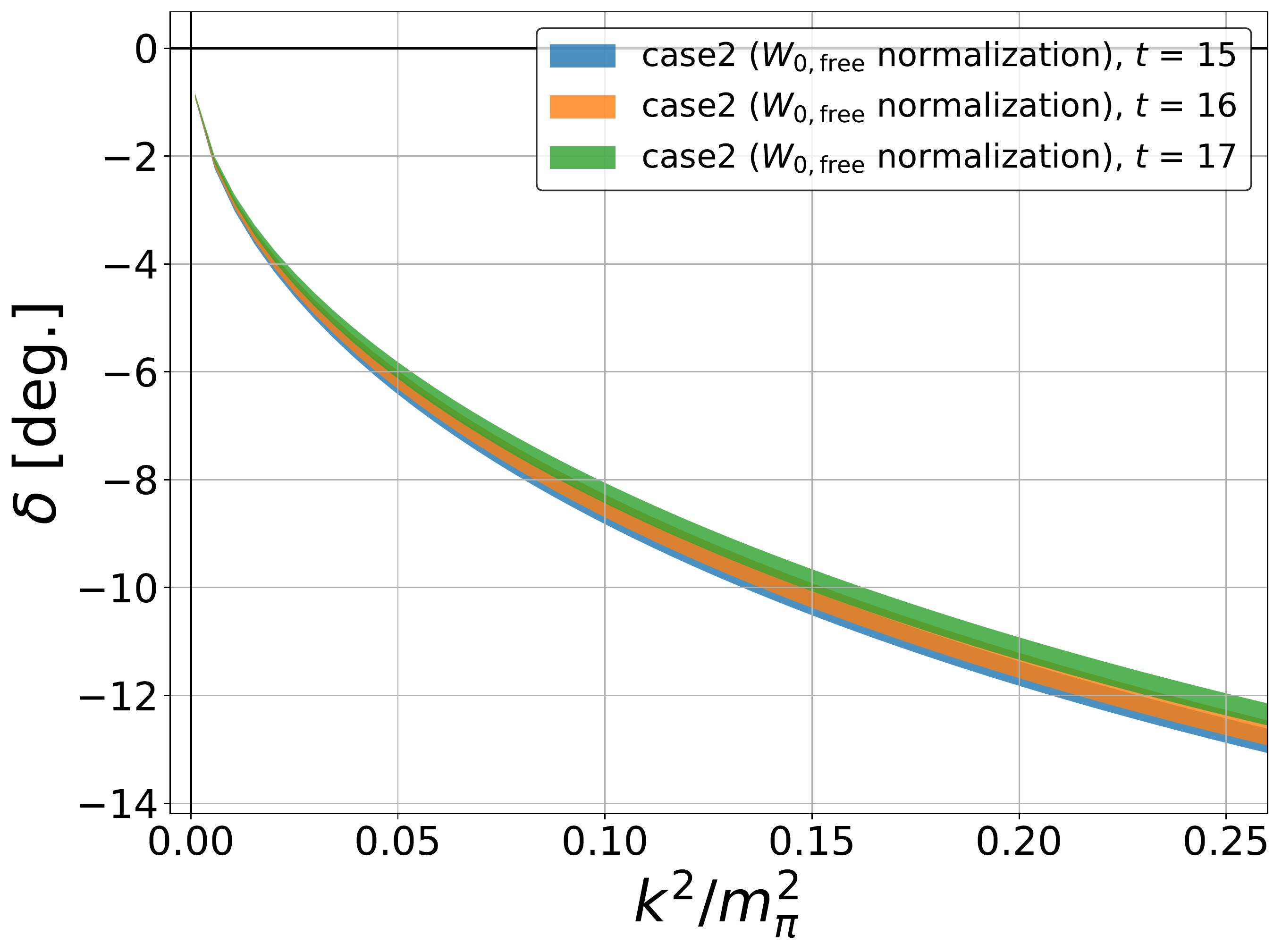}
    \end{minipage}
  \end{tabular}
  \caption{
	The $X^4$ dependence of scattering phase shifts using $W_{0,{\rm free}}$ normalization in the case 1 (Left) and the case 2 (Right). A very similar  $X^4$ dependence is observed in the case of $2m$ normalization.
  }
  \label{fig:appx:phaseshift_tdep}
\end{figure}

We next discuss the $X^4$ dependence.
Figures~\ref{fig:appx:pot_tdep_case1} (case 1) and \ref{fig:appx:pot_tdep_case2} (case 2) show effective LO potentials at $X^4 = 16 \pm 1$
with non-zero total moment using $W_{0,{\rm free}}$ normalization (Left) and $2m$ normalization (Right).
While potentials at different $X^4$ are statistically consistent with each other,
statistical fluctuations of central values slightly affect fit results of potentials.
As a result, scattering phase shifts also show a weak dependence on $X_4$, as seen in Fig.~\ref{fig:appx:phaseshift_tdep}
for the $W_{0, {\rm free}}$ normalization.
We thus include the $X_4$ dependence in  our estimation of systematic errors.

%Statistical correlation between data at different $X^4$ may cause this behavior.

\subsection{Final estimation of uncertainties}

\begin{figure}[htbp]
  \begin{tabular}{cc}
    \begin{minipage}{0.5\hsize}
        \centering
        \includegraphics[width=0.95\hsize,clip]{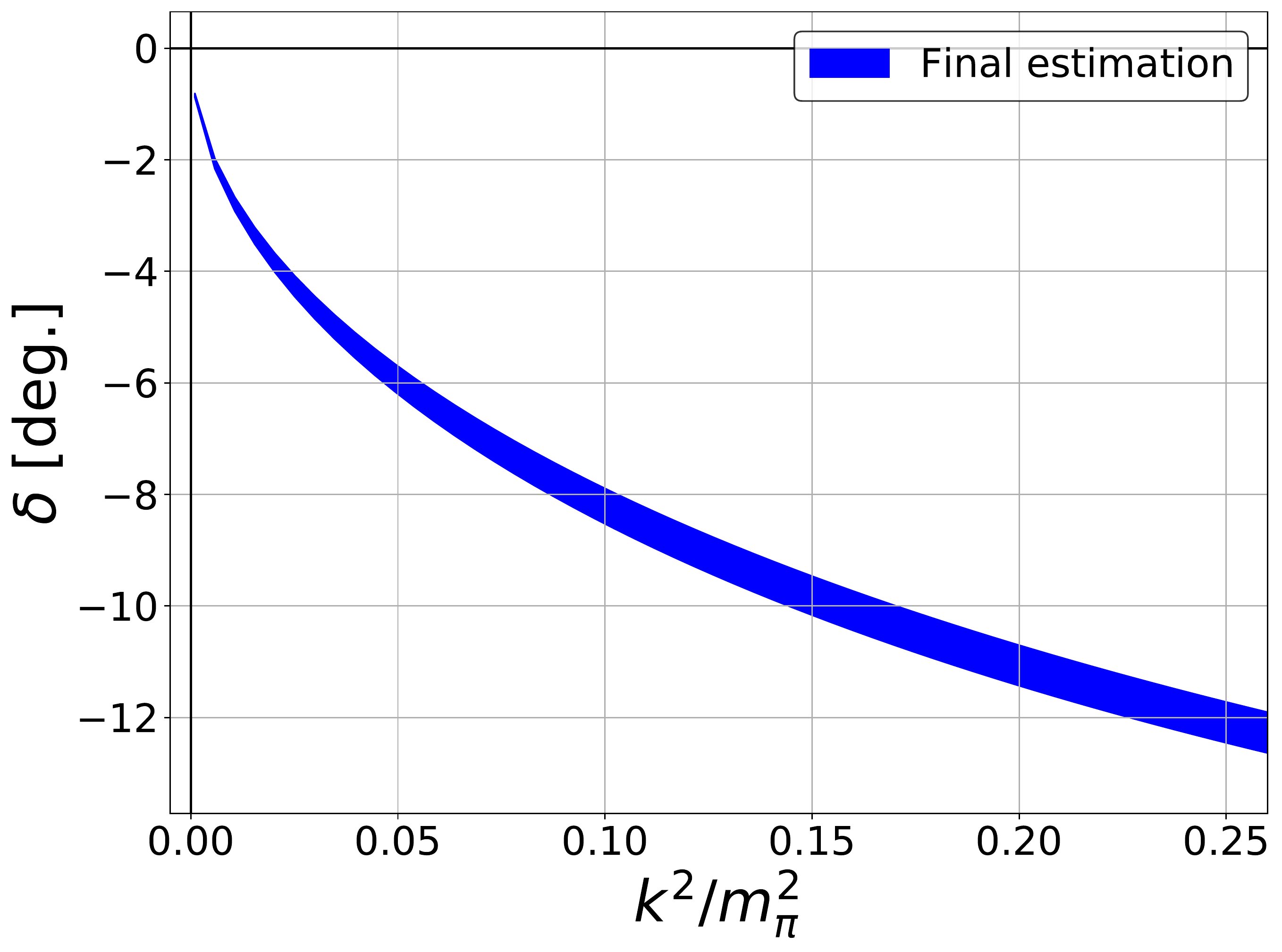}
    \end{minipage}
    &
    \begin{minipage}{0.5\hsize}
        \centering
        \includegraphics[width=0.95\hsize,clip]{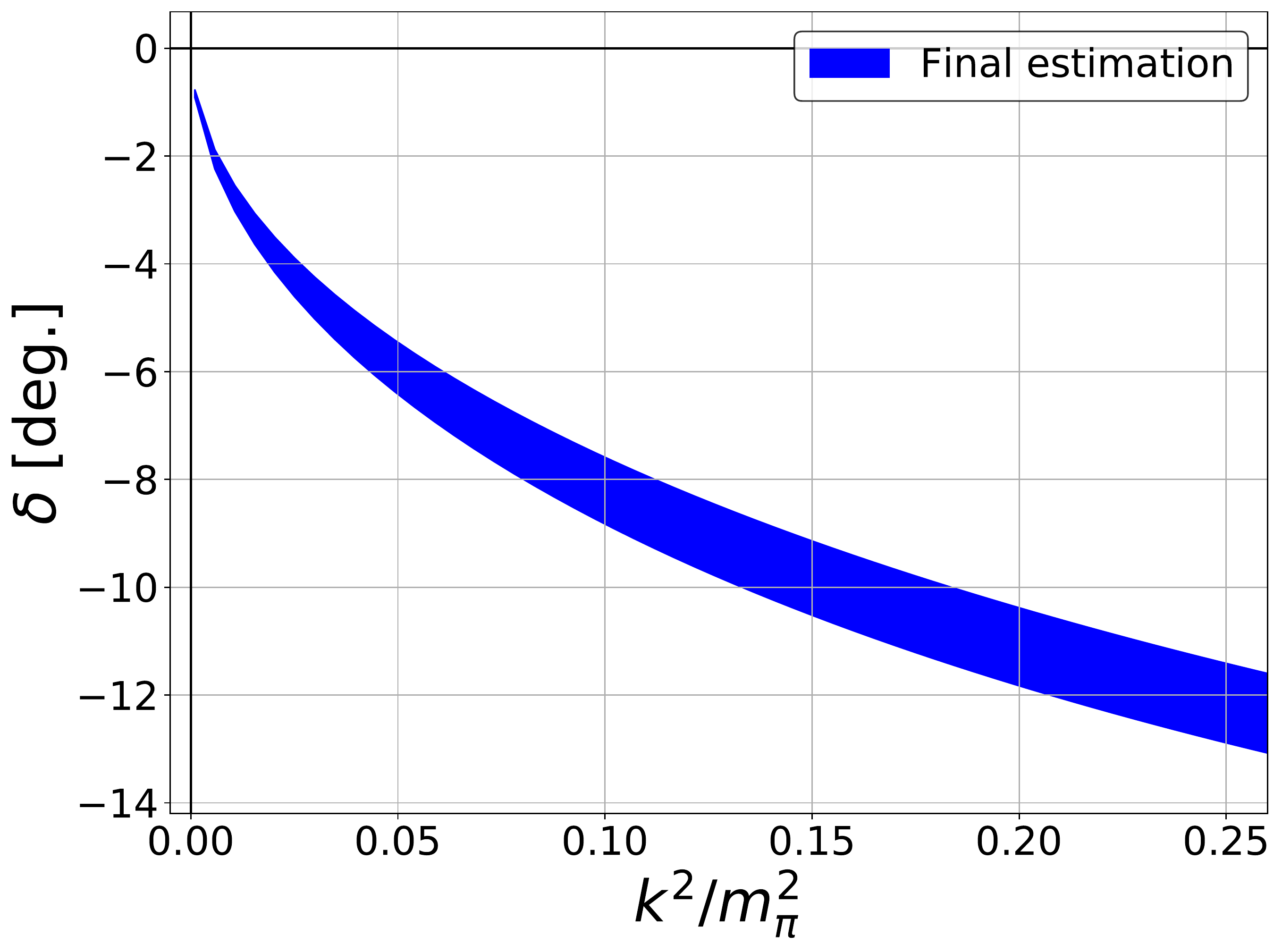}
    \end{minipage}
  \end{tabular}
  \caption{
  Scattering phase shifts  with the final estimation of uncertainties in the case 1 (Left) and the case 2 (Right). Color bands include both statistical and systematic errors.
  }
  \label{fig:appx:phase_sys}
\end{figure}

Let us present our final estimation of systematic uncertainties, including
both normalization dependence and $X^4$ dependence of scattering phase shifts.
We estimate systematic uncertainties from  differences between maximum and minimum of all data at a given energy.
Figure~\ref{fig:appx:phase_sys} shows scattering phase shifts as a function of $k^2/m_\pi^2$
with the final estimation of systematic uncertainties, where color bands include both statistical and systematic errors.
In the main text, we discuss consistency among different extractions of scattering phase shifts, taking these systematic uncertainties into account.

%%%%% Appendix end %%%%%

%%%%% ref %%%%%
%\newpage
\bibliography{ref}
\bibliographystyle{apsrev4-1.bst}

%%%%% ref end %%%%%

\end{document}